\documentclass[12pt,a4paper]{article}

\usepackage{mathtools}
\usepackage{booktabs}
\usepackage{epsfig}
\usepackage{multirow}
\usepackage{lineno}  
\usepackage{graphicx}  
\usepackage{xspace} 
\usepackage{color}
\usepackage{colortbl}
\usepackage{amsmath} 

\usepackage{amssymb}
\usepackage{amsfonts}
\usepackage{upgreek} 

\usepackage{hyperref}    
\usepackage[all]{hypcap} 

\usepackage{cite}
\usepackage{mciteplus}

\usepackage{ifthen} 
\newboolean{pdflatex}
\setboolean{pdflatex}{true} 

\newboolean{articletitles}
\setboolean{articletitles}{true} 

\newboolean{uprightparticles}
\setboolean{uprightparticles}{true} 

\long\def\symbolfootnote[#1]#2{\begingroup%
\def\thefootnote{\fnsymbol{footnote}}\footnote[#1]{#2}\endgroup}

\newcommand{\upsmm}     {\decay{\PUpsilon}{\mumu}}
\newcommand{\ups}       {\PUpsilon}

\newcommand{\ones}      {\ensuremath{\PUpsilon(1\mathrm{S})}\xspace}
\newcommand{\twos}      {\ensuremath{\PUpsilon(2\mathrm{S})}\xspace}
\newcommand{\threes}    {\ensuremath{\PUpsilon(3\mathrm{S})}\xspace}

\newcommand{\onesmm}    {\decay{\ones}{\mumu}}
\newcommand{\twosmm}    {\decay{\twos}{\mumu}}
\newcommand{\threesmm}  {\decay{\threes}{\mumu}}

\newcommand{\bit}{\begin{itemize}}
\newcommand{\bce}{\begin{center}}
\newcommand{\eit}{\end{itemize}}
\newcommand{\ece}{\end{center}}

\usepackage{microtype}
\usepackage{lineno}  
\usepackage{xspace} 
\usepackage{caption} 

\usepackage{graphicx}  
\usepackage{color}
\usepackage{colortbl}
\graphicspath{{./figs/}} 

\usepackage{amsmath} 
\usepackage{amssymb}
\usepackage{amsfonts}
\usepackage{upgreek} 

\newcommand*\patchAmsMathEnvironmentForLineno[1]{%
\expandafter\let\csname old#1\expandafter\endcsname\csname #1\endcsname
\expandafter\let\csname oldend#1\expandafter\endcsname\csname
end#1\endcsname
 \renewenvironment{#1}%
   {\linenomath\csname old#1\endcsname}%
   {\csname oldend#1\endcsname\endlinenomath}%
}
\newcommand*\patchBothAmsMathEnvironmentsForLineno[1]{%
  \patchAmsMathEnvironmentForLineno{#1}%
  \patchAmsMathEnvironmentForLineno{#1*}%
}
\AtBeginDocument{%
\patchBothAmsMathEnvironmentsForLineno{equation}%
\patchBothAmsMathEnvironmentsForLineno{align}%
\patchBothAmsMathEnvironmentsForLineno{flalign}%
\patchBothAmsMathEnvironmentsForLineno{alignat}%
\patchBothAmsMathEnvironmentsForLineno{gather}%
\patchBothAmsMathEnvironmentsForLineno{multline}%
}

\usepackage{hyperref}    
\usepackage[all]{hypcap} 

\def\lhcb {\mbox{LHCb}\xspace}
\def\ux85 {\mbox{UX85}\xspace}

\ifthenelse{\boolean{uprightparticles}}%
{

 \def\Pmu         {\ensuremath{\upmu}\xspace}

 \def\Pchi        {\ensuremath{\upchi}\xspace}                 
 \def\Ppsi        {\ensuremath{\uppsi}\xspace}

 \def\PDelta      {\ensuremath{\Delta}\xspace}                 
 \def\PXi      {\ensuremath{\Xi}\xspace}                 
 \def\PLambda      {\ensuremath{\Lambda}\xspace}                 
 \def\PSigma      {\ensuremath{\Sigma}\xspace}                 
 \def\POmega      {\ensuremath{\Omega}\xspace}                 
 \def\PUpsilon      {\ensuremath{\Upsilon}\xspace}                 
 
 \def\PB      {\ensuremath{\mathrm{B}}\xspace}                 
                  
 \def\PD      {\ensuremath{\mathrm{D}}\xspace}

 \def\PJ      {\ensuremath{\mathrm{J}}\xspace}                 
 \def\PK      {\ensuremath{\mathrm{K}}\xspace}

 \def\Pb      {\ensuremath{\mathrm{b}}\xspace}                 
 \def\Pc      {\ensuremath{\mathrm{c}}\xspace}

 \def\Pi      {\ensuremath{\mathrm{i}}\xspace}

 \def\Pp      {\ensuremath{\mathrm{p}}\xspace}

}
{

 \def\Pmu         {\ensuremath{\mu}\xspace}

 \def\Pchi        {\ensuremath{\chi}\xspace}                 
 \def\Ppsi        {\ensuremath{\psi}\xspace}                 
                  
 \mathchardef\PDelta="7101
 \mathchardef\PXi="7104
 \mathchardef\PLambda="7103
 \mathchardef\PSigma="7106
 \mathchardef\POmega="710A
 \mathchardef\PUpsilon="7107
                  
 \def\PB      {\ensuremath{B}\xspace}                 
                  
 \def\PD      {\ensuremath{D}\xspace}

 \def\PJ      {\ensuremath{J}\xspace}                 
 \def\PK      {\ensuremath{K}\xspace}

 \def\Pb      {\ensuremath{b}\xspace}                 
 \def\Pc      {\ensuremath{c}\xspace}

 \def\Pi      {\ensuremath{i}\xspace}

 \def\Pp      {\ensuremath{p}\xspace}

}

\def\mumu       {\ensuremath{\Pmu^+\Pmu^-}\xspace}

\def\cquark    {\ensuremath{\Pc}\xspace}

\def\bquark    {\ensuremath{\Pb}\xspace}
\def\bquarkbar {\ensuremath{\overline \bquark}\xspace}
\def\bbbar     {\ensuremath{\bquark\bquarkbar}\xspace}

\def\kaon  {\ensuremath{\PK}\xspace}
  \def\Kbar  {\kern 0.2em\overline{\kern -0.2em \PK}{}\xspace}

\def\Kz    {\ensuremath{\kaon^0}\xspace}
\def\Kzb   {\ensuremath{\Kbar^0}\xspace}
\def\KzKzb {\ensuremath{\Kz \kern -0.16em \Kzb}\xspace}
\def\Kp    {\ensuremath{\kaon^+}\xspace}
\def\Km    {\ensuremath{\kaon^-}\xspace}

\def\KpKm  {\ensuremath{\Kp \kern -0.16em \Km}\xspace}

  \def\Dbar    {\kern 0.2em\overline{\kern -0.2em \PD}{}\xspace}
\def\D       {\ensuremath{\PD}\xspace}

\def\Dz      {\ensuremath{\D^0}\xspace}
\def\Dzb     {\ensuremath{\Dbar^0}\xspace}
\def\DzDzb   {\ensuremath{\Dz {\kern -0.16em \Dzb}}\xspace}
\def\Dp      {\ensuremath{\D^+}\xspace}
\def\Dm      {\ensuremath{\D^-}\xspace}

\def\DpDm    {\ensuremath{\Dp {\kern -0.16em \Dm}}\xspace}

  \def\Bbar    {\kern 0.18em\overline{\kern -0.18em \PB}{}\xspace}

\def\jpsi     {\ensuremath{{\PJ\mskip -3mu/\mskip -2mu\Ppsi\mskip 2mu}}\xspace}

  \def\Y#1S{\ensuremath{\PUpsilon{(#1S)}}\xspace}

\def\proton      {\ensuremath{\Pp}\xspace}
\def\antiproton  {\ensuremath{\overline \proton}\xspace}

\def\Lbar {\ensuremath{\kern 0.1em\overline{\kern -0.1em\PLambda}}\xspace}

\def\BF         {{\ensuremath{\cal B}\xspace}}

\def\BR         {\BF}
\newcommand{\decay}[2]{\ensuremath{#1\!\to #2}\xspace}         

\def\to                 {\ensuremath{\rightarrow}\xspace}

\def\AT#1     {\ensuremath{A_{\mathrm{T}}^{#1}}\xspace}           

\def\C#1      {\ensuremath{\mathcal{C}_{#1}}\xspace}                       
\def\Cp#1     {\ensuremath{\mathcal{C}_{#1}^{'}}\xspace}                    
\def\Ceff#1   {\ensuremath{\mathcal{C}_{#1}^{\mathrm{(eff)}}}\xspace}        
\def\Cpeff#1  {\ensuremath{\mathcal{C}_{#1}^{'\mathrm{(eff)}}}\xspace}       
\def\Ope#1    {\ensuremath{\mathcal{O}_{#1}}\xspace}                       
\def\Opep#1   {\ensuremath{\mathcal{O}_{#1}^{'}}\xspace}                    

\newcommand{\tev}{\ensuremath{\mathrm{\,Te\kern -0.1em V}}\xspace}
\newcommand{\gev}{\ensuremath{\mathrm{\,Ge\kern -0.1em V}}\xspace}
\newcommand{\mev}{\ensuremath{\mathrm{\,Me\kern -0.1em V}}\xspace}
\newcommand{\kev}{\ensuremath{\mathrm{\,ke\kern -0.1em V}}\xspace}
\newcommand{\ev}{\ensuremath{\mathrm{\,e\kern -0.1em V}}\xspace}
\newcommand{\gevc}{\ensuremath{{\mathrm{\,Ge\kern -0.1em V\!/}c}}\xspace}
\newcommand{\mevc}{\ensuremath{{\mathrm{\,Me\kern -0.1em V\!/}c}}\xspace}
\newcommand{\gevcc}{\ensuremath{{\mathrm{\,Ge\kern -0.1em V\!/}c^2}}\xspace}
\newcommand{\gevgevcccc}{\ensuremath{{\mathrm{\,Ge\kern -0.1em V^2\!/}c^4}}\xspace}
\newcommand{\mevcc}{\ensuremath{{\mathrm{\,Me\kern -0.1em V\!/}c^2}}\xspace}

\def\mum  {\ensuremath{\,\upmu\rm m}\xspace}

\def\nb {\ensuremath{\rm \,nb}\xspace}

\def\invpb {\ensuremath{\mbox{\,pb}^{-1}}\xspace}

\newcommand{\chisq}{\ensuremath{\chi^2}\xspace}

\def\gsim{{~\raise.15em\hbox{$>$}\kern-.85em
          \lower.35em\hbox{$\sim$}~}\xspace}
\def\lsim{{~\raise.15em\hbox{$<$}\kern-.85em
          \lower.35em\hbox{$\sim$}~}\xspace}

\def\sPlot{\mbox{\em sPlot}}

\def\sqs   {\ensuremath{\protect\sqrt{s}}\xspace}

\def\pt         {\mbox{$p_{\rm T}$}\xspace}

\newcommand{\lum} {\ensuremath{\mathcal{L}}\xspace}

\def\evtgen     {\mbox{\textsc{EvtGen}}\xspace}

\def\geant      {\mbox{\textsc{Geant4}}\xspace}

\def\photos     {\mbox{\textsc{Photos}}\xspace}

\def\tell1  {TELL1\xspace}
\def\ukl1   {UKL1\xspace}

\usepackage{cite} 
\usepackage{mciteplus}

\usepackage{rotating} 
\usepackage{graphpap} 
\usepackage{multirow} 
\usepackage{mathrsfs} 
\usepackage{pifont}   
\usepackage{cancel}   

\begin{document}


\renewcommand{\thefootnote}{\fnsymbol{footnote}}
\setcounter{footnote}{1}


\begin{titlepage}
\pagenumbering{roman}

\vspace*{-1.5cm}
\centerline{\large EUROPEAN ORGANIZATION FOR NUCLEAR RESEARCH (CERN)}
\vspace*{1.5cm}
\hspace*{-0.5cm}
\begin{tabular*}{\linewidth}{lc@{\extracolsep{\fill}}r}
\ifthenelse{\boolean{pdflatex}}
{\vspace*{-2.7cm}\mbox{\!\!\!\includegraphics[width=.14\textwidth]{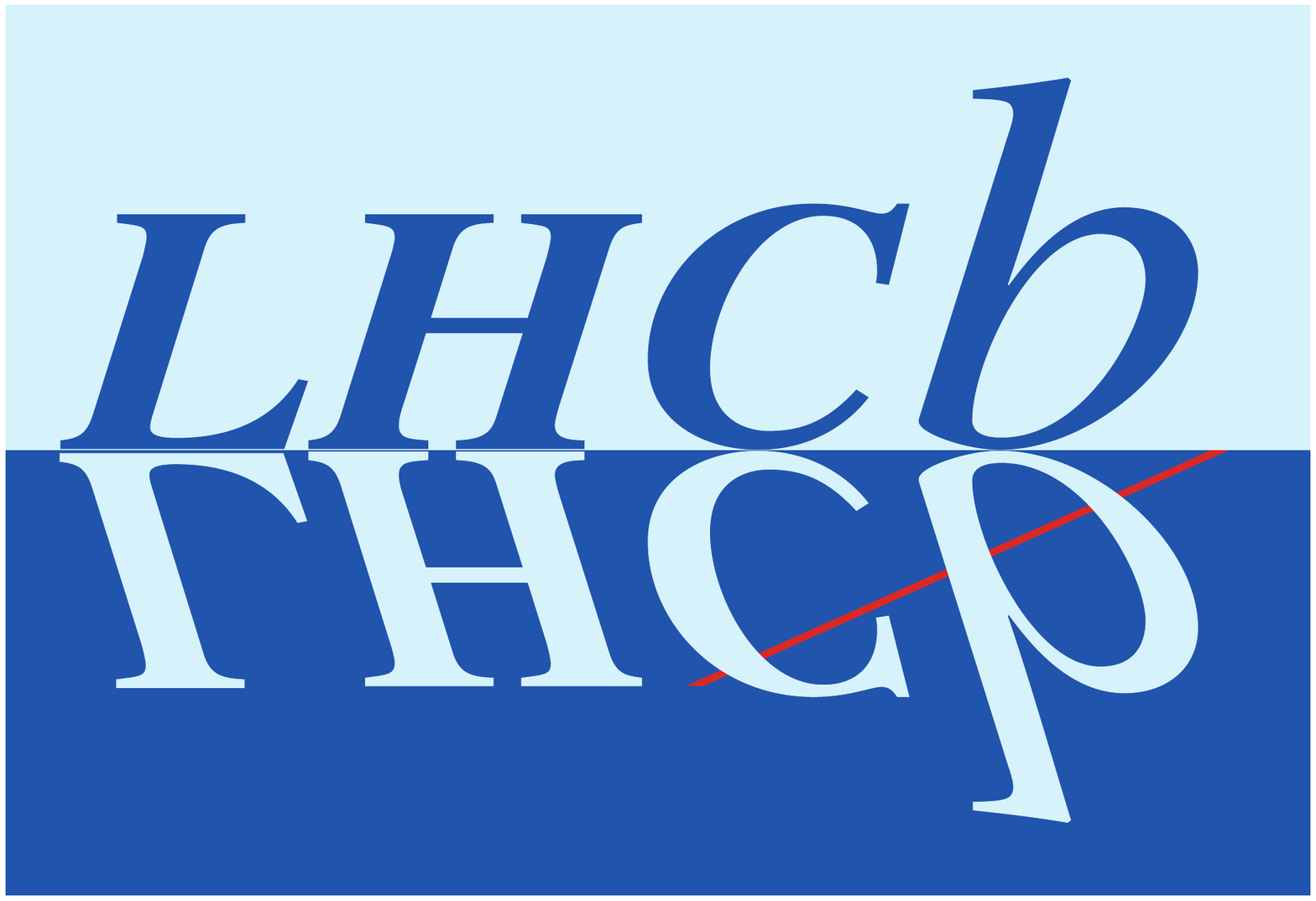}} & &}%
{\vspace*{-1.2cm}\mbox{\!\!\!\includegraphics[width=.12\textwidth]{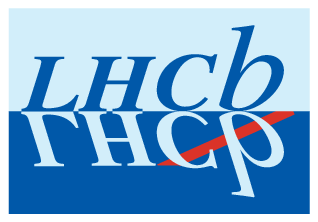}} & &}%
\\
 & & CERN-PH-EP-2014-016 \\  
 & & LHCb-PAPER-2013-066 \\  
 & & 25 March 2014 \\ 
\end{tabular*}

\vspace*{2.0cm}

{\bf\boldmath\huge
\begin{center}
Measurement of $\ups$~production in $\proton\proton$~collisions at $\sqs=2.76\tev$
\end{center}
}

\vspace*{1.0cm}

\begin{center}
The LHCb collaboration\symbolfootnote[1]{Authors are listed on the following pages.}
\end{center}

\vspace{\fill}

\begin{abstract}
  \noindent
  The production of $\ones$, $\twos$ and $\threes$ mesons decaying
  into the dimuon final state is studied with the LHCb detector
  using a data sample corresponding to an~integrated luminosity 
  of~$3.3\invpb$ collected
  in proton-proton collisions at a~centre-of-mass energy of $\sqrt{s}=2.76$~TeV.
  The~differential production cross-sections times dimuon branching fractions 
  are measured as functions of the $\ups$ transverse momentum and rapidity,
  over the ranges $p_{\rm T}<15$~GeV/$c$ and $2.0<y<4.5$.
  The~total cross-sections in this kinematic region, assuming unpolarised production,  
  are measured to be 
  \begin{eqnarray*}
  \upsigma \left( \proton\proton \rightarrow \ones   \mathrm{X} \right) \times \BR\left(\onesmm   \right)  & = & 1.111 \pm 0.043  \pm 0.044\nb, \\
  \upsigma \left( \proton\proton \rightarrow \twos   \mathrm{X} \right) \times \BR\left(\twosmm   \right)  & = & 0.264 \pm 0.023  \pm 0.011\nb,  \\ 
  \upsigma \left( \proton\proton \rightarrow \threes \mathrm{X} \right) \times \BR\left(\threesmm \right)  & = & 0.159 \pm 0.020  \pm 0.007\nb,
  \end{eqnarray*}
  where the first uncertainty is statistical and the second systematic.

\end{abstract}

\vspace*{1.5cm}

\begin{center}
 Submitted to Eur.~Phys.~J.~C 
\end{center}

\vspace{\fill}

{\footnotesize 
\centerline{\copyright~CERN on behalf of the \lhcb collaboration, license \href{http://creativecommons.org/licenses/by/3.0/}{CC-BY-3.0}.}}
\vspace*{2mm}

\end{titlepage}


\newpage
\setcounter{page}{2}
\mbox{~}
\newpage

\centerline{\large\bf LHCb collaboration}
\begin{flushleft}
\small
R.~Aaij$^{41}$, 
B.~Adeva$^{37}$, 
M.~Adinolfi$^{46}$, 
A.~Affolder$^{52}$, 
Z.~Ajaltouni$^{5}$, 
J.~Albrecht$^{9}$, 
F.~Alessio$^{38}$, 
M.~Alexander$^{51}$, 
S.~Ali$^{41}$, 
G.~Alkhazov$^{30}$
% ...

{\footnotesize \it
% ...
$ ^{5}$Clermont Universit\'{e}, Universit\'{e} Blaise Pascal, CNRS/IN2P3, LPC, Clermont-Ferrand, France\\
% ...
$ ^{9}$Fakult\"{a}t Physik, Technische Universit\"{a}t Dortmund, Dortmund, Germany\\
% ...
$ ^{30}$Petersburg Nuclear Physics Institute (PNPI), Gatchina, Russia\\
% ...
$ ^{37}$Universidad de Santiago de Compostela, Santiago de Compostela, Spain\\
$ ^{38}$European Organization for Nuclear Research (CERN), Geneva, Switzerland\\
% ...
$ ^{41}$Nikhef National Institute for Subatomic Physics, Amsterdam, The Netherlands\\
% ...
$ ^{46}$H.H. Wills Physics Laboratory, University of Bristol, Bristol, United Kingdom\\
% ...
$ ^{51}$School of Physics and Astronomy, University of Glasgow, Glasgow, United Kingdom\\
$ ^{52}$Oliver Lodge Laboratory, University of Liverpool, Liverpool, United Kingdom\\
% ...
}
\end{flushleft}

\cleardoublepage


\renewcommand{\thefootnote}{\arabic{footnote}}
\setcounter{footnote}{0}

\cleardoublepage


\pagestyle{plain} 
\setcounter{page}{1}
\pagenumbering{arabic}


\section{Introduction}
\label{sec:intro}

Studies of the production of heavy quark-antiquark bound systems, such as 
the \bbbar~states \ones, \twos and \threes (indicated generically 
as \ups~in the following) in hadron-hadron interactions probe the dynamics 
of the colliding partons and provide a unique insight into quantum chromodynamics (QCD). 
The total production cross-sections and spin configurations of these heavy quarkonium states
are currently not reproduced by the~theoretical models. 
These include the colour singlet model~\cite{Kartvelishvili:1978id,CSM,Berger:1980ni,CSM1,CSM2}, 
recently improved by adding higher-order contributions~\cite{Campbell:2007ws,Artoisenet:2008fc}, 
the colour-evaporation model~\cite{CEM}, 
and the non-perturbative colour octet mechanism~\cite{COM1,COM2,COM3},
which is investigated in the framework of non-relativistic QCD.
The first complete next-to-leading order calculation 
of \ups production properties~\cite{Gong:2013qka},
based on the non-relativistic QCD factorisation scheme, 
provides a good description of the measured differential cross-sections 
at large transverse momentum, \pt, 
but overestimates the data at low \pt. 

The production of \ups~mesons in proton-proton\,($\proton\proton$) collisions 
occurs either directly in parton scattering or via feed-down from 
the decay of heavier prompt bottomonium states, 
like $\Pchi_{\bquark}$~\cite{Aad:2011ih,Abazov:2012gh,LHCB-PAPER-2012-015,LHCb-CONF-2012-020}, 
or higher-mass $\ups$ states.
The latter source complicates the theoretical description 
of bottomonium production~\cite{Likhoded:2012hw,Wang:2012is}.

The Large Hadron Collider provides a unique possibility to study 
bottomonium and charmonium hadroproduction in $\proton\proton$ 
interactions at different collision energies and discriminate 
between various theoretical approaches.
This study presents the first measurement of the inclusive production 
cross-sections of the three considered $\ups$~mesons in $\proton\proton$~collisions 
at a~centre-of-mass energy of $\sqs=2.76\tev$.
The measurements are performed as functions of the~$\ups$~transverse momentum 
and rapidity, $y$, separately in six bins of $\pt$ in the~range $\pt<15\gevc$ 
and five bins of $y$ in the~range $2.0<y<4.5$.
The results are reported as~products of the cross-sections and the branching  
fractions of \ups~mesons into the dimuon final state.  
This analysis is complementary to those performed by the ATLAS~\cite{Aad:2012dlq}, CMS~\cite{Chatrchyan:2013yna} 
and LHCb~\cite{LHCb-PAPER-2011-036,LHCb-PAPER-2013-016} collaborations 
and allows studies of the~$\ups$~production cross-section at forward rapidities 
as a~function of the~centre-of-mass energy.

\section{Detector and data sample}
\label{sec:lhcb}

The \lhcb detector~\cite{Alves:2008zz} is a single-arm forward
spectrometer covering the \mbox{pseudorapidity} range $2<\eta<5$,
designed for the study of particles containing \bquark~or \cquark~quarks. 
The~detector includes a high-precision tracking system
consisting of a silicon-strip vertex detector surrounding the $\proton\proton$
interaction region, a large-area silicon-strip detector located
upstream of a dipole magnet with a bending power of about
$4{\rm\,Tm}$, and three stations of silicon-strip detectors and straw
drift tubes placed downstream. The combined tracking system provides 
a momentum measurement with relative uncertainty that varies from 0.4\% 
at 5\gevc to 0.6\% at 100\gevc, and impact parameter resolution of 20\mum for
tracks with large transverse momentum. Different types of charged hadrons 
are distinguished by information from two ring-imaging Cherenkov detectors~\cite{LHCb-DP-2012-003}. 
Photon, electron and hadron candidates are identified by a calorimeter system 
consisting of scintillating-pad and preshower detectors, an electromagnetic
calorimeter and a hadronic calorimeter. Muons are identified by a
system composed of alternating layers of iron and multiwire
proportional chambers~\cite{LHCb-DP-2012-002}.

The analysis is carried out using a sample of data corresponding 
to an integrated luminosity of $3.3\invpb$ collected 
in $\proton\proton$ collisions at $\sqs=2.76\tev$. 
Events of interest are preselected by a trigger consisting 
of a hardware stage, based on information from 
the~calorimeter and muon systems, followed by 
a software stage, which applies a~full event reconstruction.  
The presence of two muon candidates with the~product of their \pt larger 
than 1.68~$($GeV$/c$$)^{2}$ is required in the hardware trigger.
At the software stage, the~events are required to contain 
two well reconstructed tracks with hits in the muon system, 
having total and transverse momenta greater than $6\gevc$ and $0.5\gevc$, respectively. 
The~selected muon candidates are further required to originate from a common vertex 
and have an invariant mass larger than $4.7\gevcc$.

To determine the acceptance, reconstruction and trigger efficiencies, 
fully simulated signal samples are reweighted to reproduce
the multiplicity distributions for reconstructed primary vertices, 
tracks and hits in the detector observed in the data. 
The simulation is performed using the \lhcb configuration~\cite{LHCb-PROC-2010-056} 
of the {\sc Pythia}~6.4 event generator~\cite{Sjostrand:2006za}.
Here, decays of hadronic particles are described by \evtgen~\cite{Lange:2001uf}
in which final-state photons are generated using \photos~\cite{Golonka:2005pn}.
The interaction of the generated particles with the~detector and its
response are implemented using the \geant toolkit~\cite{Allison:2006ve, Agostinelli:2002hh}
as described in Ref.~\cite{LHCb-PROC-2011-006}.

\section{Signal selection and cross-section determination}
\label{sec:analysis}

The selection strategy used in the previous LHCb studies 
on \ups production~\cite{LHCb-PAPER-2011-036,LHCb-PAPER-2013-016} 
is applied here. It includes selection criteria that 
ensure good quality track and vertex reconstruction.
In addition, the muon candidates 
are required to have $p>10\gevc$ and $\pt>1\gevc$. 
To~further reduce background contamination, a set of additional 
requirements is employed in this analysis. 
It consists of tightened criteria
on track quality~\cite{LHCb-DP-2013-002}, 
muon identification~\cite{LHCb-DP-2013-001}
and a~good quality of a~global fit 
of the~dimuon vertex with a~primary vertex constraint~\cite{Hulsbergen:2005pu}.   

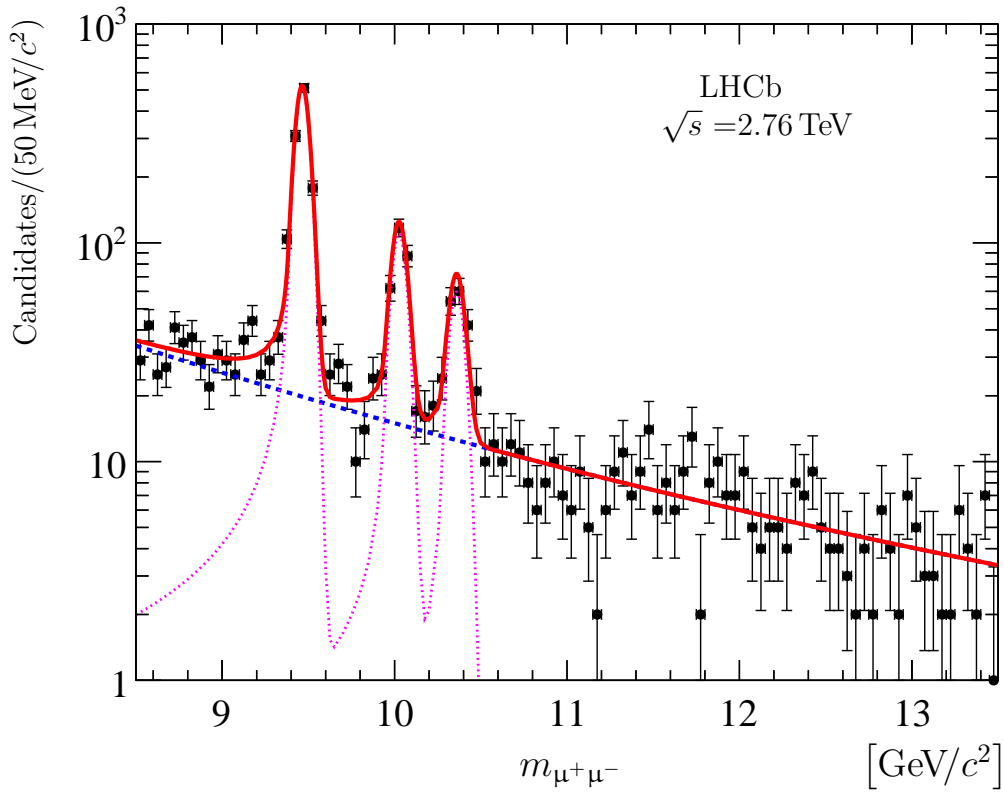
\begin{figure}[t]
  \setlength{\unitlength}{1mm}
  \centering
  \begin{picture}(140,110)
    \put(0,0){ 
      \includegraphics*[width=140mm,height=110mm,%
      ]{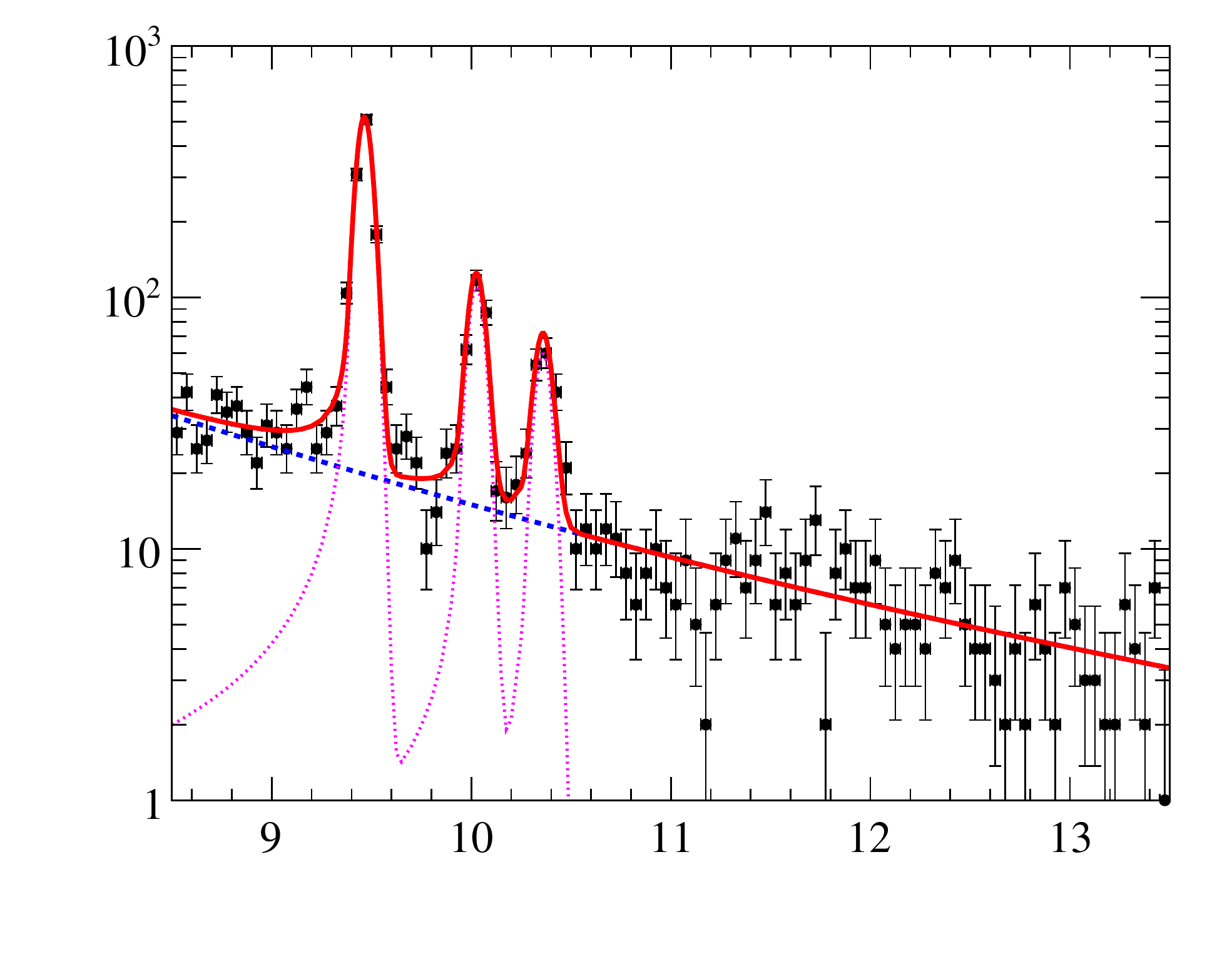}
    }
    \put( 95,95){LHCb}
    \put( 90,90){\sqs=2.76\tev }
    \put( 70,5){ \large $m_{\mumu}$  }
    \put(115,5){ \large $\left[\!\gevcc\right]$  }
    \put(3,62) { \begin{sideways}Candidates/(50\mevcc)\end{sideways}}
  \end{picture}
  \caption { \small
    Invariant mass distribution of selected $\upsmm$ candidates
    with $\pt<15\gevc$ and $2.0<y<4.5$.
    The result of the fit described in the text is illustrated with a red solid line, 
    while the signal and background components are shown 
    with magenta dotted and blue dashed lines, respectively.
    The three peaks correspond to the $\ones$, $\twos$ and $\threes$ mesons 
    (from left to right).
  }
  \label{fig:Ymass}
\end{figure}

The invariant mass distribution of the selected $\upsmm$ candidates is shown 
in Fig.~\ref{fig:Ymass} for the full kinematic range. 
The distribution is described by a function similar 
to the one used in the previous studies on \ups~production~\cite{LHCb-PAPER-2011-036,LHCb-PAPER-2013-016}.
It models the signal component using the sum of three  
Crystal Ball functions~\cite{Skwarnicki:1986xj}, 
one for each of the $\ones$, $\twos$ and $\threes$ signals, 
and includes an exponential component
to account for combinatorial background.
The position and width of the Crystal Ball function describing 
the $\ones$ meson are allowed to vary, while the mass differences between \ups states 
are fixed to their known values~\cite{PDG2012} along with parameters describing 
the radiative tail, as determined from simulation studies.  
The widths of the $\twos$ and $\threes$ peaks are constrained to the value of 
the width of the $\ones$ signal scaled by the ratio of their masses 
to the $\ones$ mass. 
In total, five parameters are extracted from the fit 
for the signal component: the yields of $\ones$, $\twos$ and $\threes$ states, 
the $\ones$ mass resolution and its peak position.
The latter is found to be consistent with 
the known mass of the $\ones$ meson~\cite{PDG2012}, while 
reasonable agreement is observed 
between the data and simulation for the $\ones$ mass resolution.

The \ups~production cross-sections are measured 
separately in six bins of \pt and five bins of $y$ since 
the limited amount of data does not allow a measurement of double differential cross-sections. 
For a given \pt or $y$ bin, the differential cross-section for the~inclusive 
\ups~production of the three different states decaying into the dimuon final state 
is determined as
\begin{subequations}
\begin{equation}
\dfrac{ {\mathrm{d}}\upsigma \left( \proton\proton \rightarrow \ups {\mathrm{X}} \right)}
      {\mathrm{d}\pt}\times \mathcal{B}\left( \upsmm \right)   =  
\dfrac{ N^{\mathrm{corr}}_{\ups}}{\lum \times \Delta \pt } \;,  
\end{equation}
\begin{equation}
\dfrac{ {\mathrm {d}}\upsigma \left( \proton\proton \rightarrow \ups {\mathrm{X}}\right)}{\mathrm{d}y}  \times \mathcal{B}\left( \upsmm \right)  =  
\dfrac{ N^{\mathrm{corr}}_{\ups}}{\lum \times \Delta y } \;,
\end{equation}
\label{eq:sigma}
\end{subequations}
where $N^{\mathrm{corr}}_{\ups}$ is the efficiency-corrected yield
of $\upsmm$~decays, \lum~stands for the~integrated luminosity and
$\Delta \pt\,(\Delta y)$~denotes the $\pt\,(y)$~bin size. 
For the mass fits in individual \pt and $y$ bins, 
the~\ones peak position is fixed to the~value 
obtained from the~fit for the~full kinematic range, 
while the~\ones~mass resolution is parameterised 
with a~function of \pt~and $y$~using simulation. 
The total observed signal yields and their statistical uncertainties for $\ones$, 
$\twos$ and $\threes$ mesons 
obtained by summation over 
$\pt\,(y)$~bins are
$1139\pm37\,(1145\pm37)$, 
$271\pm20\,(270\pm20)$ and 
$158\pm16\,(156\pm16)$, respectively.
These~results are in good agreement with the total signal yields obtained 
from the fit to the~reconstructed dimuon invariant mass for the full kinematic range.

Based on the mass fit results in individual bins, 
the efficiency-corrected yield for each kinematic region is determined as 
\begin{equation}
N^{\mathrm{corr}}_{\ups} = \sum_{i} \dfrac{ w^{\ups}_{i} }{ \varepsilon^{\mathrm{tot}}_{i} } \;,
\label{eq:ncorr}
\end{equation}
where $w^{\ups}_{i}$~is a signal weight factor, 
$\varepsilon^{\mathrm{tot}}_{i}$~is the total signal event efficiency 
and the sum runs over all candidates $i$.
The $w^{\ups}_{i}$ factor accounts for the~background subtraction and  
is obtained from the~fit using the~\sPlot~technique~\cite{Pivk:2004ty}. 
The total signal event efficiency is calculated for 
each $\upsmm$~candidate as 
\begin{equation}
\varepsilon^{\mathrm{tot}} = 
\varepsilon^{\mathrm{acc}}     \times 
\varepsilon^{\mathrm{rec}} \times 
\varepsilon^{\mathrm{trg}}     \times 
\varepsilon^{\Pmu\mathrm{ID}} \;,  
\label{eq:effic}
\end{equation}
where 
$\varepsilon^{\mathrm{acc}}$~is the~detector acceptance,
$\varepsilon^{\mathrm{rec}}$~is the reconstruction and selection efficiency, 
$\varepsilon^{\mathrm{trg}}$~is the~trigger efficiency 
and  $\varepsilon^{\Pmu\mathrm{ID}}$~is the efficiency of muon identification.  
The efficiencies  $\varepsilon^{\mathrm{acc}}$, 
$\varepsilon^{\mathrm{rec}}$ and 
$\varepsilon^{\mathrm{trg}}$~are determined
using simulation and further corrected 
using data-driven techniques to account for 
small differences in muon reconstruction 
efficiency between data and simulation~\cite{LHCb-DP-2013-001,LHCb-DP-2013-002,LHCb-PAPER-2010-001}. 
The efficiency~$\varepsilon^{\Pmu\mathrm{ID}}$ is measured directly from 
data using a~tag-and-probe method on
a~large sample of $\jpsi\to\mumu$~decays.
The~total efficiency-corrected signal yields 
obtained by summation over 
$\pt\,(y)$~bins for $\ones$, 
$\twos$ and $\threes$ mesons are
$3678\pm144\,(3684\pm143)$, 
$875\pm76\,(869\pm75)$ and 
$527\pm65\,(515\pm64)$, respectively. 

The integrated luminosity of the~data sample 
is estimated with the beam-gas imaging 
method~\cite{FerroLuzzi:2005em,LHCb-PAPER-2011-015,Hopchev:2011ht,Hopchev,Barschel}. 
It is based on the beam currents and the measurements of 
the~angles, 
offsets and transverse profiles of the two colliding bunches, 
which is achieved by reconstructing beam-gas interaction vertices. 

\section{Systematic uncertainties}
\label{sec:syst}
 
Previous LHCb studies of \ups~production~\cite{LHCb-PAPER-2011-036,LHCb-PAPER-2013-016}
showed that the signal efficiency depends on the initial polarisation of \ups mesons.
This property was measured in $\proton\proton$~collisions at $\sqs=7\tev$ by the CMS 
collaboration  at central rapidities and large \pt and was found to be small~\cite{Chatrchyan:2012woa}.
Polarisation of other vector quarkonium states, 
such as \jpsi~and $\Ppsi\mathrm{(2S)}$~mesons 
was studied in $\proton\proton$~collisions at $\sqs=7\tev$ by the 
LHCb~\cite{LHCb-PAPER-2013-008,LHCb-PAPER-2013-067} and ALICE~\cite{Abelev:2011md} 
collaborations and was also found to be small. 
This analysis is performed assuming 
zero polarisation of \ups~mesons and 
no corresponding systematic uncertainty is assigned.

\begin{table}[t]
  \centering
  \caption{ \small
    Relative systematic uncertainties (in $\%$) affecting 
    the \ups production cross-section measurements in the~full kinematic region.
    The total uncertainties are obtained 
    by adding the~individual effects in quadrature.
  } \label{tab:syst_summary}
  \vspace*{3mm}
  \begin{tabular*}{0.95\textwidth}{@{\hspace{3mm}}l@{\extracolsep{\fill}}cccc@{\hspace{3mm}}}
    Source                              &   \ones & \twos & \threes  \\
    \hline  
    Luminosity                          &   2.3   & 2.3  &  2.3  \\
    Fit model and range                 &   0.5   & 1.0  &  2.3  \\
    Data-simulation agreement           & & & \\ 
    ~~~Radiative tails                  &   1.0   & 1.0  &  1.0  \\
    ~~~Multiplicity reweighting         &   0.6   & 0.4  &  2.0  \\   
    ~~~Efficiency corrections           &   0.7   & 1.0  &  1.0  \\    
    ~~~Track reconstruction             &   $2\times0.4$ 
                                        &   $2\times0.4$ 
                                        &   $2\times0.4$         \\ 
    ~~~Selection variables              &   1.0   & 1.0  &  1.0  \\
    ~~~Trigger                          &   2.0   & 2.0  &  2.0  \\ 
   \hline 
   Total                                &   3.6   & 3.7  &  4.7  
  \end{tabular*}   
\end{table}

The systematic uncertainties affecting the \ups cross-section measurements presented 
in this paper are summarised in Table~\ref{tab:syst_summary}. 
These uncertainties are strongly correlated between bins. 
The largest contribution arises  from the absolute luminosity scale, 
which is determined with a~2.3\% uncertainty. 
It is dominated by the vertex resolution of beam-gas interactions and detector 
alignment~\cite{Barschel}.    

The influence of the signal extraction technique 
is studied by varying the fit range and 
the signal and background parameterisations used in the fit model. 
The fits are also performed with floating mass and resolution of the \ones~peak 
and  without constraints for the~\twos and \threes masses. 
The spread of the extracted signal yields between these scenarios 
is taken as the corresponding systematic uncertainty. 
It~ranges from 0.4 to 33\% for different $\pt\,(y)$~bins
and amounts to 0.5\%, 1.0\%  and 2.3\% for the
\ones, \twos and \threes~cross-section measurements in the~full kinematic region, respectively. 

The possible mismodeling of bremsstrahlung simulation for the radiative tail
and its effect on the signal shape was addressed in the previous 
LHCb analysis~\cite{LHCb-PAPER-2013-016}.
It leads to an~additional uncertainty of 1.0\%.

Several systematic uncertainties are related to the determination of the 
total efficiency components in Eq.~\eqref{eq:effic}.
The detector acceptance, reconstruction and selection efficiencies 
are determined using simulated samples.
These are corrected using an iterative procedure 
to match the multiplicity distributions for reconstructed 
primary vertices, tracks and hits in the detector with those 
observed in data. The systematic uncertainty associated with 
this reweighting procedure is assessed 
by varying the number of iterative steps.
It~ranges from 0.4 to 4.8\% for different $\pt\,(y)$~bins
and is found to be 0.6\%, 0.4\% and 2.0\% for 
the~\ones, \twos and \threes~cross-section measurements 
in the~full kinematic region, respectively.

The $\varepsilon^{\mathrm{rec}}$ efficiency is corrected using data-driven 
techniques for a small difference in the~muon reconstruction efficiency 
between data and simulation~\cite{LHCb-DP-2013-001,LHCb-DP-2013-002}.
The~$\varepsilon^{\Pmu\mathrm{ID}}$~efficiency
is determined from data using alternative methods, based on a tag-and-probe approach 
on a large sample of $\jpsi\to\mumu$ decays.
The difference between these methods is taken as the corresponding systematic uncertainty.
It is combined with the~uncertainties associated with the~correction factors 
discussed above and propagated to the~\ups~cross-section measurements using 
400~pseudo-experiments. The resulting uncertainty 
ranges from 1.0 to 13\% for different $\pt\,(y)$~bins and
amounts to 0.7\%, 1.0\% and 1.0\% 
for the \ones, \twos and \threes~cross-section measurements 
in the~full kinematic region, respectively. 

To account for differences between the~actual tracking efficiency and 
that estimated with simulation using data-driven techniques~\cite{LHCb-DP-2013-002,LHCb-PAPER-2010-001}, 
a systematic uncertainty of 0.4\% is assigned per track. 

Good agreement between the data and reweighted simulation is observed for all selection 
variables used in this analysis, in particular for the~\chisq~of 
the~dimuon vertex fit  and the~\chisq~of the~global~fit~\cite{Hulsbergen:2005pu}. 
The discrepancies do not 
exceed 1.0\%, which is conservatively taken as 
a~systematic uncertainty to account for the disagreement 
between the data and simulation.

The systematic uncertainty associated with the trigger requirements 
is assessed by studying the performance of the dimuon trigger, 
described in Sect.~\ref{sec:lhcb}, for events selected using 
the~single muon high-\pt~trigger~\cite{LHCb-DP-2012-004}.
The fractions of signal \ones~events selected using both trigger 
requirements are compared for the data and simulation in bins 
of dimuon \pt, and a systematic uncertainty of 2.0\% is assigned.

\section{Results}
\label{sec:results}

The integrated \ups production cross-sections times dimuon branching fractions
in the~kinematic region $\pt<15\gevc$ and $2.0<y<4.5$ are measured to be
\begin{eqnarray*}
  \upsigma \left( \proton\proton \rightarrow \ones   \mathrm{X} \right) \times \BR\left(\onesmm   \right)  & = & 1.111 \pm 0.043  \pm 0.044\nb, \\
  \upsigma \left( \proton\proton \rightarrow \twos   \mathrm{X} \right) \times \BR\left(\twosmm   \right)  & = & 0.264 \pm 0.023  \pm 0.011\nb,  \\ 
  \upsigma \left( \proton\proton \rightarrow \threes \mathrm{X} \right) \times \BR\left(\threesmm \right)  & = & 0.159 \pm 0.020  \pm 0.007\nb,
\end{eqnarray*}
where the first uncertainty is statistical and the second systematic. 

\begin{figure}[t!]
  \setlength{\unitlength}{1mm}
  \centering
  \begin{picture}(150,180)
    \put( 0,120){
      \includegraphics*[width=75mm,height=60mm,%
      ]{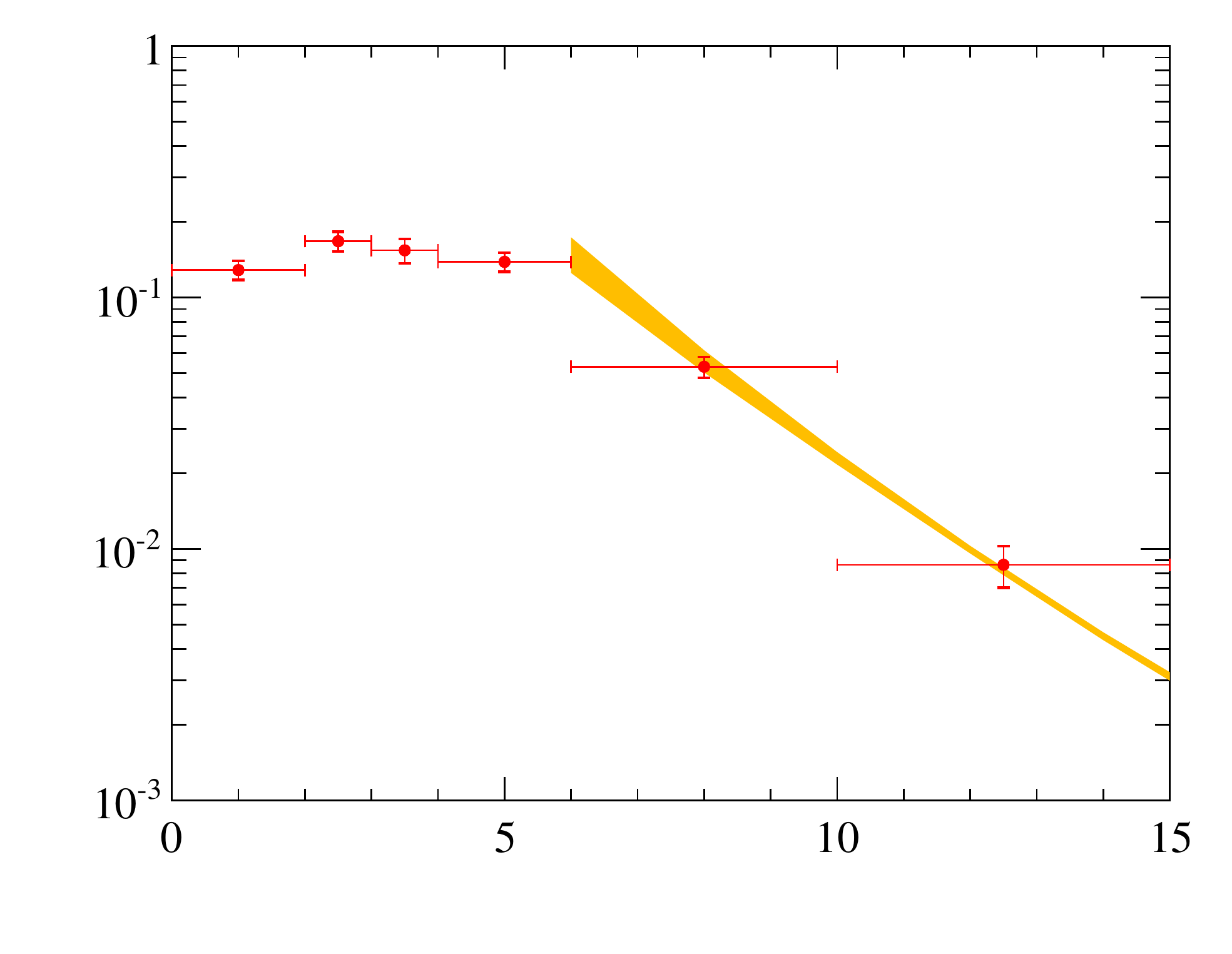}
    }
    \put(75,120){
      \includegraphics*[width=75mm,height=60mm,%
      ]{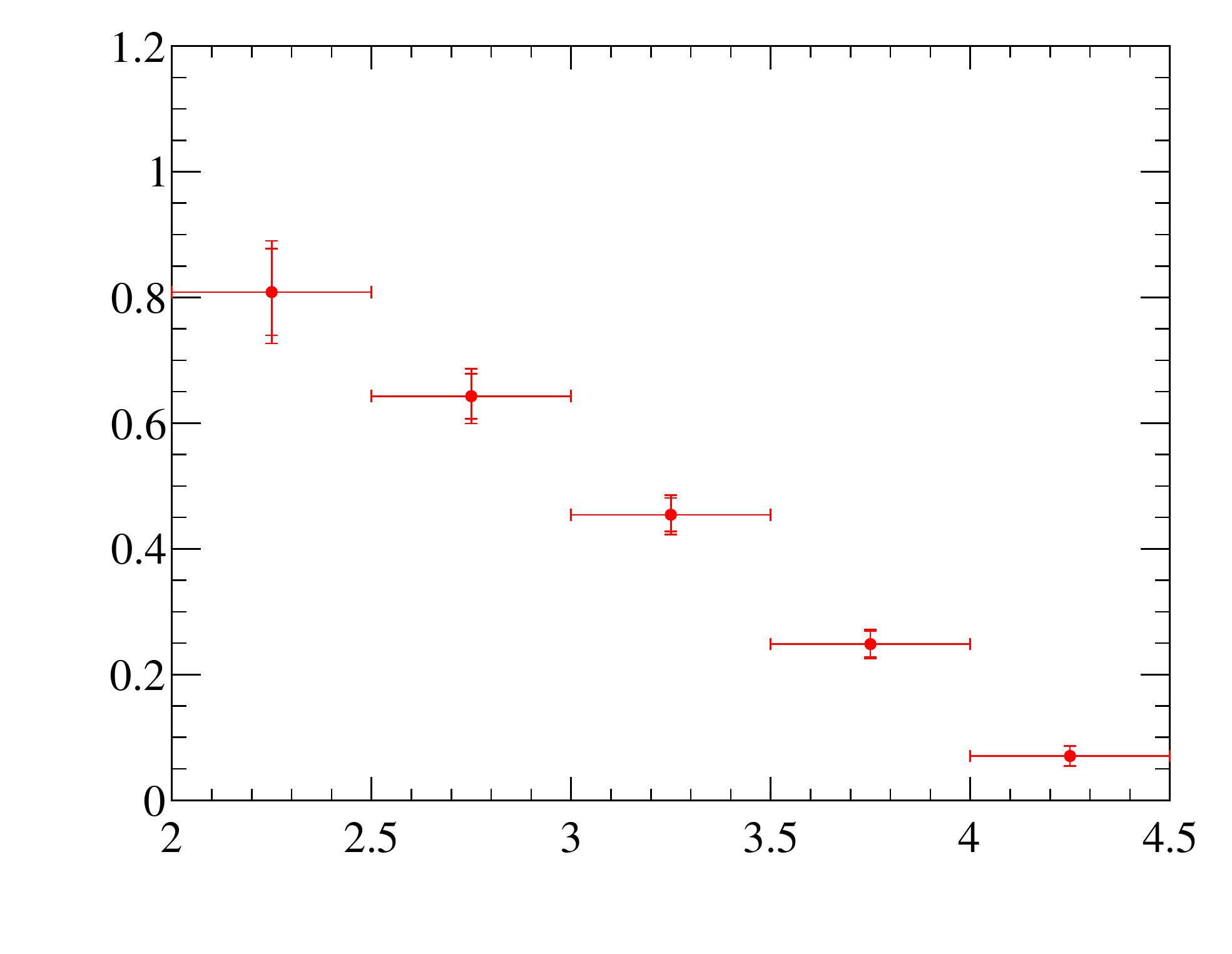}
    }
    \put( 0, 60){
      \includegraphics*[width=75mm,height=60mm,%
      ]{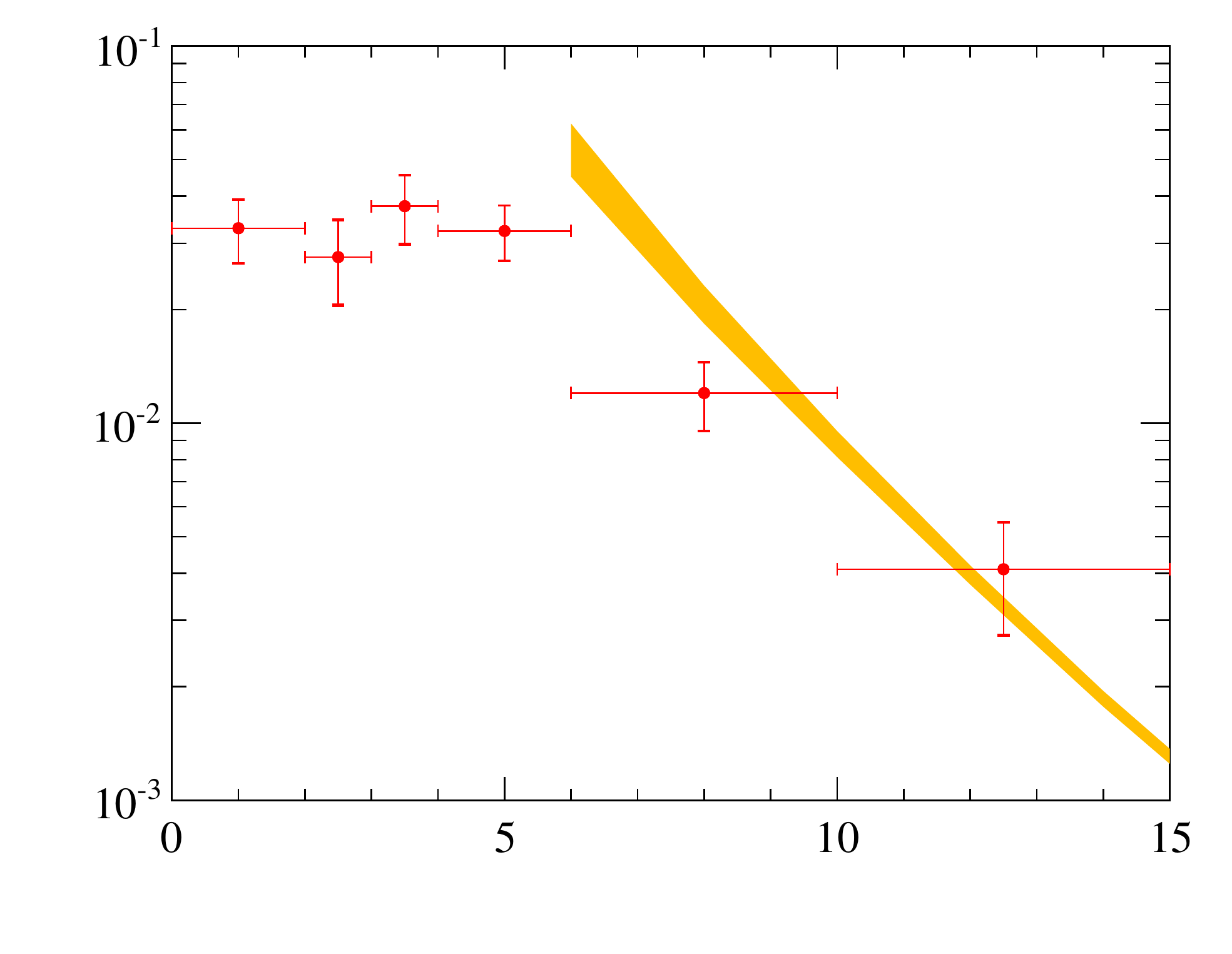}
    }
    \put(75, 60){
      \includegraphics*[width=75mm,height=60mm,%
      ]{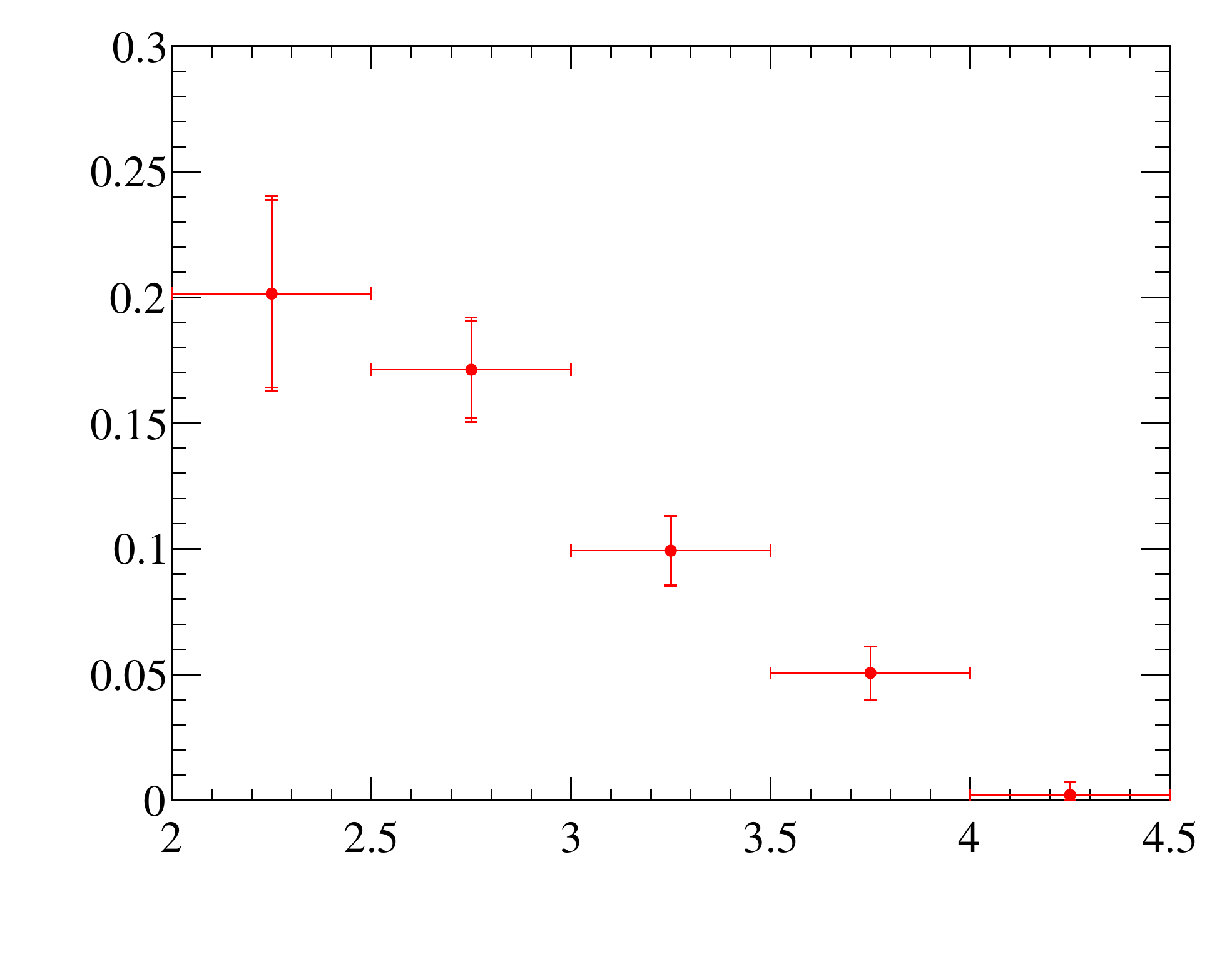}
    }
    \put( 0, 0){
      \includegraphics*[width=75mm,height=60mm,%
      ]{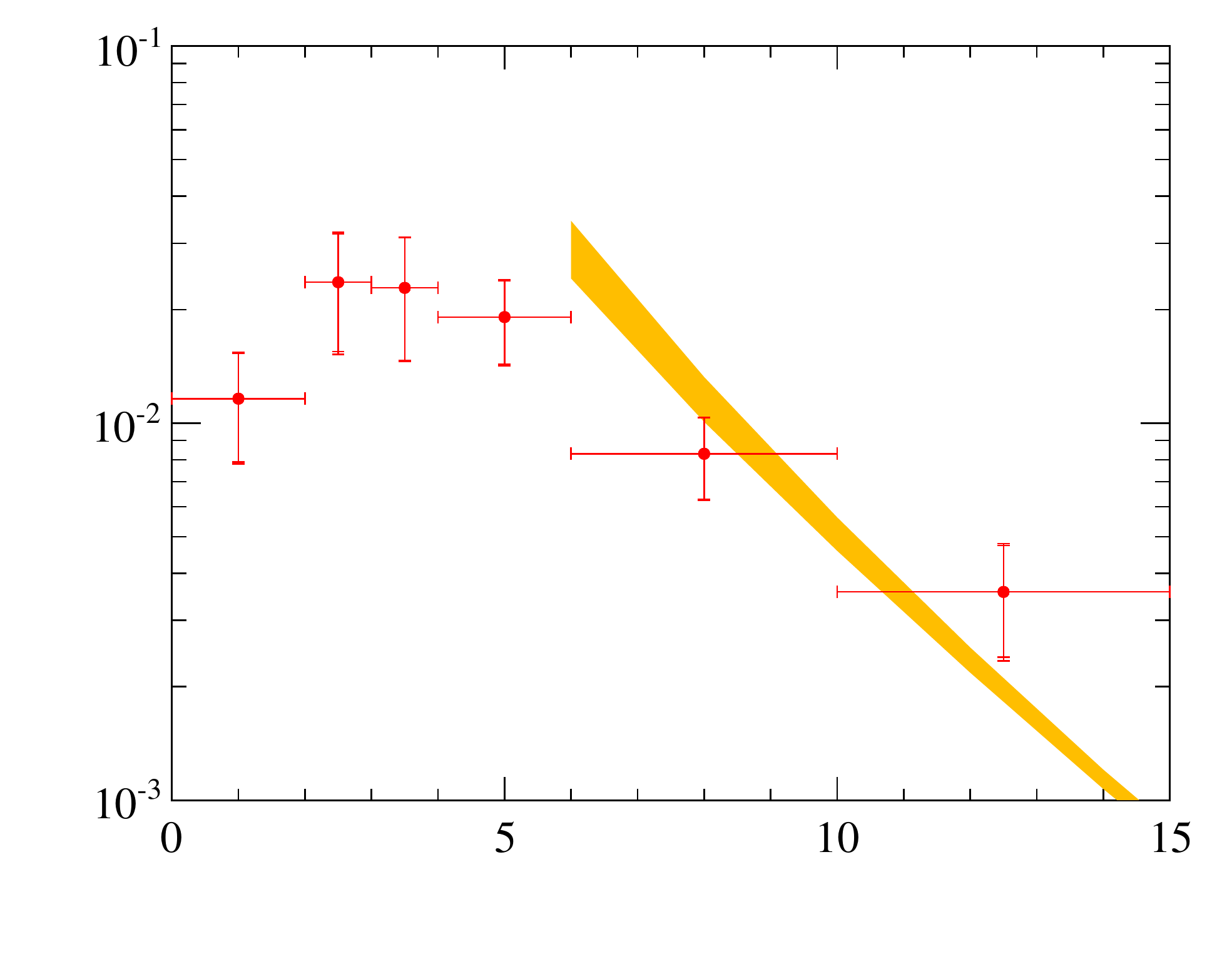}
    }
    \put(75, 0){
      \includegraphics*[width=75mm,height=60mm,%
      ]{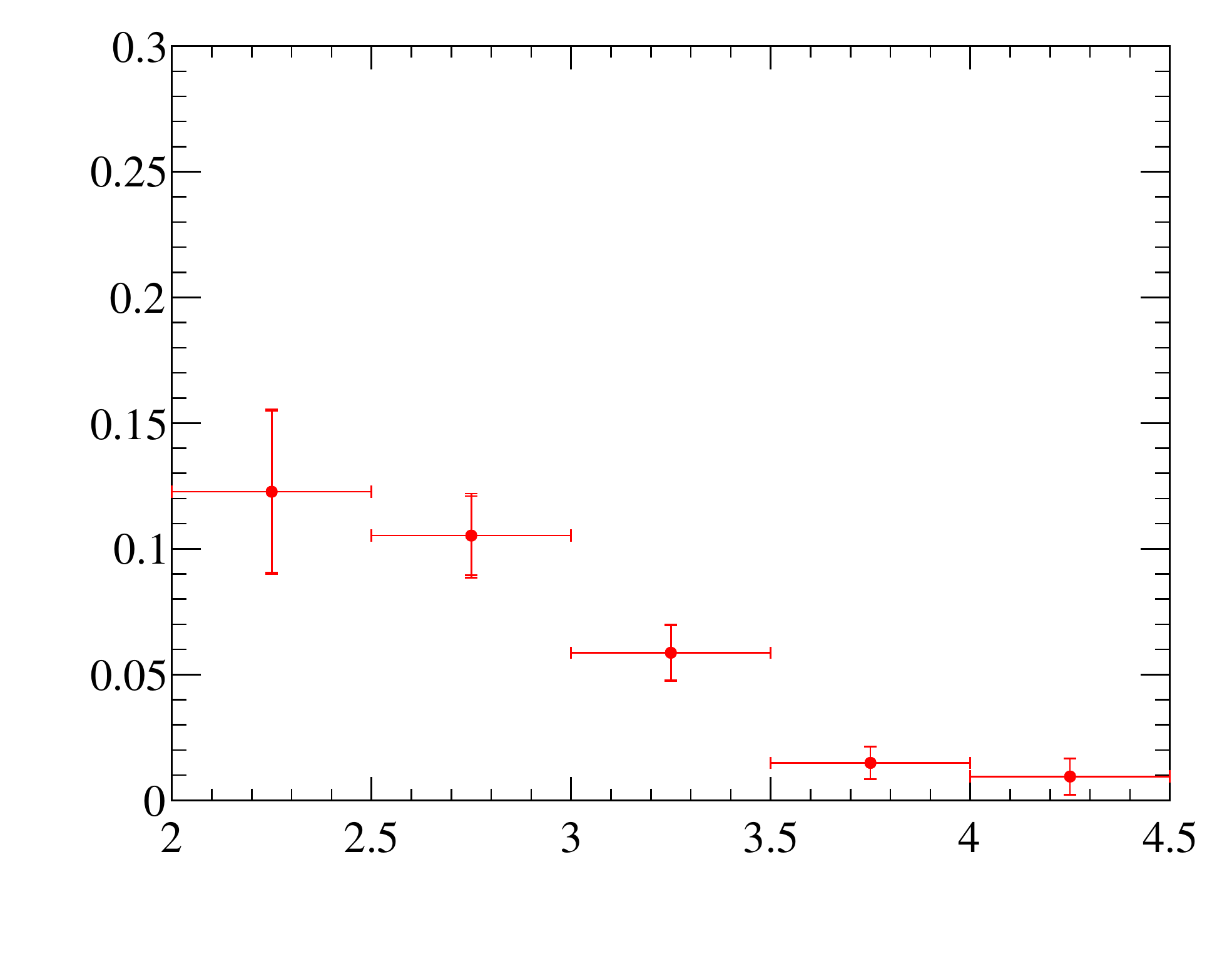}
    }
    \put( 49,170){LHCb}
    \put( 42,165){$\onesmm$}
    \put(122,170){LHCb}
    \put(115,165){$\onesmm$}
    \put( 49,110){LHCb}
    \put( 42,105){$\twosmm$}
    \put(122,110){LHCb}
    \put(115,105){$\twosmm$}
    \put( 49,50){LHCb}
    \put( 42, 45){$\threesmm$}
    \put(122,50){LHCb}
    \put(115, 45){$\threesmm$}
    \put( 38  ,121){ $\pt$  }
    \put( 57.5,121){ $\left[\!\gevc\right]$  }
    \put(112  ,121){ $y$   }
    \put( 38  , 61){ $\pt$  }
    \put( 57.5, 61){ $\left[\!\gevc\right]$  }
    \put(112  , 61){ $y$   }
    \put( 38  ,  1){ $\pt$  }
    \put( 57.5,  1){ $\left[\!\gevc\right]$  }
    \put(112  ,  1){ $y$   }
    \put(-3,130) { \small \begin{sideways} $\mathcal{B}_{\onesmm}\times\tfrac{\mathrm{d}\upsigma}{\mathrm{d}\pt}~\left[\tfrac{\mathrm{nb}}{\gevc}\right]$ \end{sideways}}
    \put(75,140) { \small \begin{sideways} $\mathcal{B}_{\onesmm}\times\tfrac{\mathrm{d}\upsigma}{\mathrm{d}y}~\left[\mathrm{nb}\right]$ \end{sideways}}
    \put(-3, 70) { \small \begin{sideways} $\mathcal{B}_{\twosmm}\times\tfrac{\mathrm{d}\upsigma}{\mathrm{d}\pt}~\left[\tfrac{\mathrm{nb}}{\gevc}\right]$ \end{sideways}}
    \put(75, 80) { \small \begin{sideways} $\mathcal{B}_{\twosmm}\times\tfrac{\mathrm{d}\upsigma}{\mathrm{d}y}~\left[\mathrm{nb}\right]$ \end{sideways}}
    \put(-3, 10) { \small \begin{sideways} $\mathcal{B}_{\threesmm}\times\tfrac{\mathrm{d}\upsigma}{\mathrm{d}\pt}~\left[\tfrac{\mathrm{nb}}{\gevc}\right]$ \end{sideways}}
    \put(75, 20) { \small \begin{sideways} $\mathcal{B}_{\threesmm}\times\tfrac{\mathrm{d}\upsigma}{\mathrm{d}y}~\left[\mathrm{nb}\right]$ \end{sideways}}
  \end{picture}
  \caption { \small
    Differential cross-sections for \ones, \twos and \threes mesons times 
    dimuon branching fractions as functions of \pt~(left) and $y$~(right).
    The~inner error bars indicate the statistical uncertainty, while the outer error bars indicate 
    the sum of statistical and systematic uncertainties in quadrature.
    The next-to-leading order non-relativistic QCD predictions~\cite{Wang:2012is}
    are shown by the~solid yellow band.
  }
  \label{fig:results_xs_1}
\end{figure}
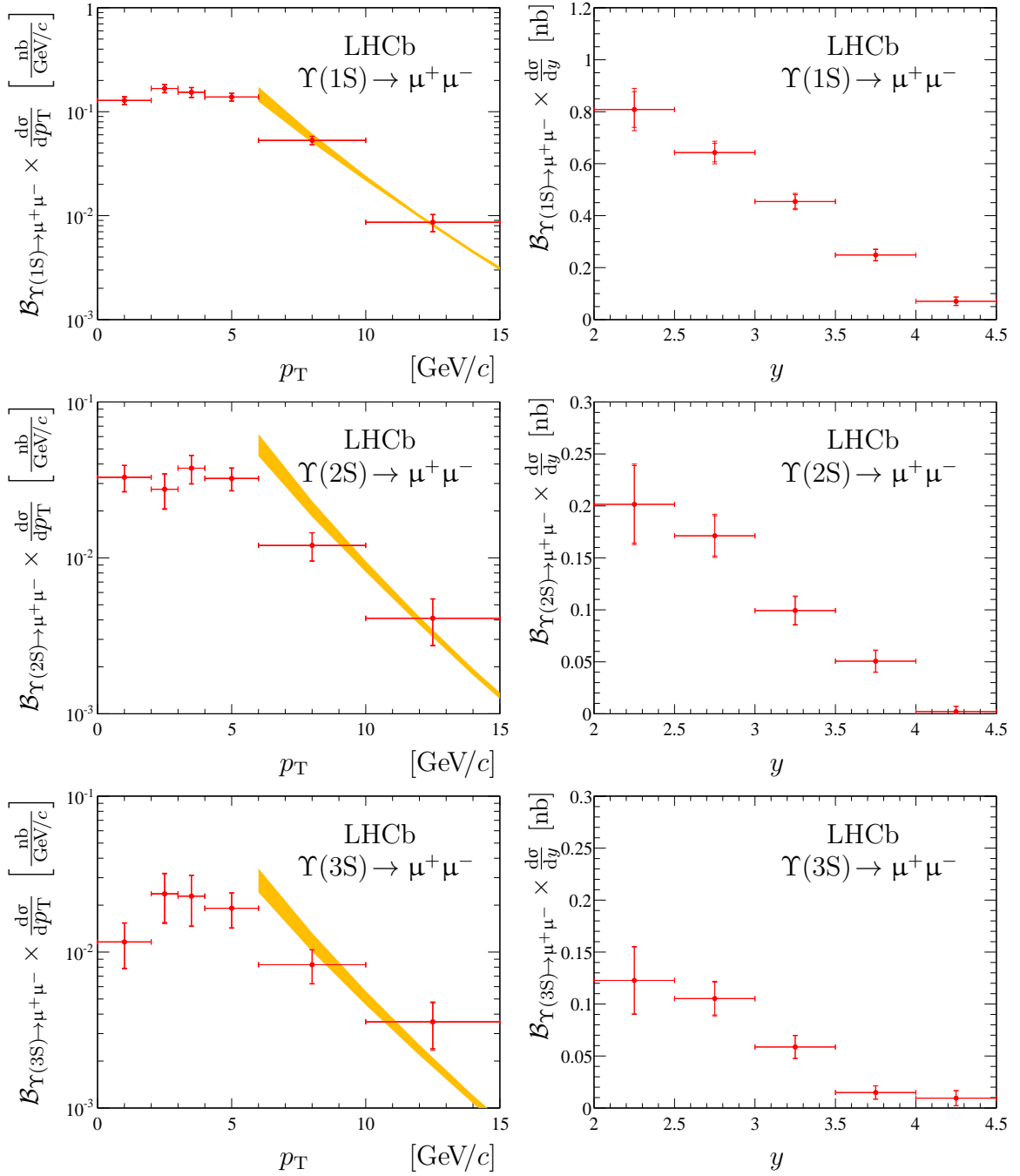

The single differential cross-sections times dimuon branching 
fractions are shown as functions of \pt and $y$ 
in Fig.~\ref{fig:results_xs_1} and summarised in Table~\ref{tab:xsect}. 
The total uncertainties of the~results are dominated 
by statistical effects in all \pt and $y$ bins.  
In addition to the data, 
Fig.~\ref{fig:results_xs_1} reports 
theoretical predictions, based on the next-to-leading order 
non-relativistic QCD calculation~\cite{Wang:2012is},  
for the \ups differential cross-sections 
in the kinematic region $6<\pt<15\gevc$ and $2.0<y<4.5$.
The long-distance matrix elements used in 
the~calculations 
are fitted to CDF~\cite{Acosta:2001gv} and 
D0~\cite{Abazov:2005yc} results  
for \ones~production in $\proton\antiproton$~collisions
at $\sqrt{s}=1.8$ and 1.96\tev. 
The predictions include the~feed-down contributions 
from higher excited S-wave and P-wave $\bquark\bquarkbar$~states. 
Good agreement between the data and predictions is found 
for all three $\Upsilon$ states.
The dependence of the \ups cross-sections on $y$ is found to be more pronounced than at 
higher collision energies~\cite{LHCb-PAPER-2011-036,LHCb-PAPER-2013-016}, which is in line 
with theoretical expectations presented for example in Ref.~\cite{Kisslinger:2012tu}.

\begin{table}[t!]
\centering
\caption{ \small 
  Cross-sections for \ones, \twos and \threes mesons times 
  dimuon branching fractions (in~\nb) in bins of \pt
  and $y$~without normalisation to the bin sizes. 
  The first uncertainty is statistical  and the second is systematic.
}
  \label{tab:xsect}
\begin{tabular*}{0.95\textwidth}{@{\hspace{3mm}}c@{\extracolsep{\fill}}ccc@{\hspace{3mm}}}
$\pt~\left[\!\gevc\right]$ & \onesmm & \twosmm & \threesmm \\  
\hline
0--2   & $0.257\pm0.021\pm0.011$ & $0.066\pm0.012\pm0.007$ & $0.023\pm0.007\pm0.002$ \\
2--3   & $0.167\pm0.014\pm0.007$ & $0.028\pm0.007\pm0.002$ & $0.024\pm0.008\pm0.002$ \\
3--4   & $0.154\pm0.016\pm0.009$ & $0.038\pm0.008\pm0.002$ & $0.023\pm0.008\pm0.001$ \\
4--6   & $0.277\pm0.023\pm0.013$ & $0.065\pm0.011\pm0.003$ & $0.038\pm0.010\pm0.002$ \\
~6--10  & $0.212\pm0.019\pm0.008$ & $0.048\pm0.010\pm0.002$ & $0.033\pm0.008\pm0.001$ \\
10--15 & $0.043\pm0.008\pm0.003$ & $0.020\pm0.007\pm0.001$ & $0.018\pm0.006\pm0.002$ \\
       &  &  &  \\
$y$    & $\onesmm$ & $\twosmm$ & $\threesmm$   \\
\hline
2.0--2.5  & $0.404\pm0.034\pm0.022$ & $0.101\pm0.019\pm0.005$ & $0.061\pm0.016\pm0.003$ \\
2.5--3.0  & $0.321\pm0.018\pm0.012$ & $0.086\pm0.010\pm0.004$ & $0.053\pm0.008\pm0.003$ \\
3.0--3.5  & $0.227\pm0.013\pm0.008$ & $0.050\pm0.007\pm0.002$ & $0.029\pm0.005\pm0.001$ \\
3.5--4.0  & $0.124\pm0.011\pm0.005$ & $0.025\pm0.005\pm0.001$ & $0.007\pm0.003\pm0.001$ \\
4.0--4.5  & $0.035\pm0.008\pm0.002$ & $0.001\pm0.003\pm0.001$ & $0.005\pm0.004\pm0.001$ \\
\end{tabular*}
\end{table}

\begin{table}[h!]
\centering
\caption{ \small 
  Ratios of the $\twos$ to $\ones$ and $\threes$ to $\ones$ cross-sections times
  dimuon branching fractions as functions of \pt and $y$.
  The first uncertainty is statistical and the second is systematic.}
\label{tab:ratio}
\begin{tabular*}{0.85\textwidth}{@{\hspace{5mm}}c@{\extracolsep{\fill}}cc@{\hspace{5mm}}}
$\pt~\left[\!\gevc\right]$ &  $\mathcal{R}^{\mathrm 2S/1S}$ & $\mathcal{R}^{\mathrm 3S/1S}$  \\ 
\hline
0--2   &  $0.257\pm0.053\pm0.009$ & $0.090\pm0.030\pm0.006$   \\
2--3   &  $0.165\pm0.044\pm0.007$ & $0.141\pm0.050\pm0.010$   \\
3--4   &  $0.244\pm0.056\pm0.007$ & $0.148\pm0.055\pm0.006$   \\
4--6   &  $0.233\pm0.043\pm0.007$ & $0.138\pm0.037\pm0.005$   \\
~6--10  &  $0.227\pm0.051\pm0.006$ & $0.157\pm0.041\pm0.004$   \\
10--15 &  $0.474\pm0.179\pm0.031$ & $0.413\pm0.155\pm0.029$   \\
       &  &    \\
$y$    &  &    \\ 
\hline
2.0--2.5  & $0.249\pm0.051\pm0.007$ & $0.152\pm0.042\pm0.006$   \\
2.5--3.0  & $0.266\pm0.033\pm0.007$ & $0.164\pm0.026\pm0.007$   \\
3.0--3.5  & $0.219\pm0.032\pm0.004$ & $0.129\pm0.025\pm0.003$   \\
3.5--4.0  & $0.204\pm0.046\pm0.004$ & $0.060\pm0.026\pm0.003$   \\
\end{tabular*}
\end{table}

\begin{figure}[t]
  \setlength{\unitlength}{1mm}
  \centering
  \begin{picture}(150,120)
    \put( 0, 60){
      \includegraphics*[width=75mm,height=60mm,%
      ]{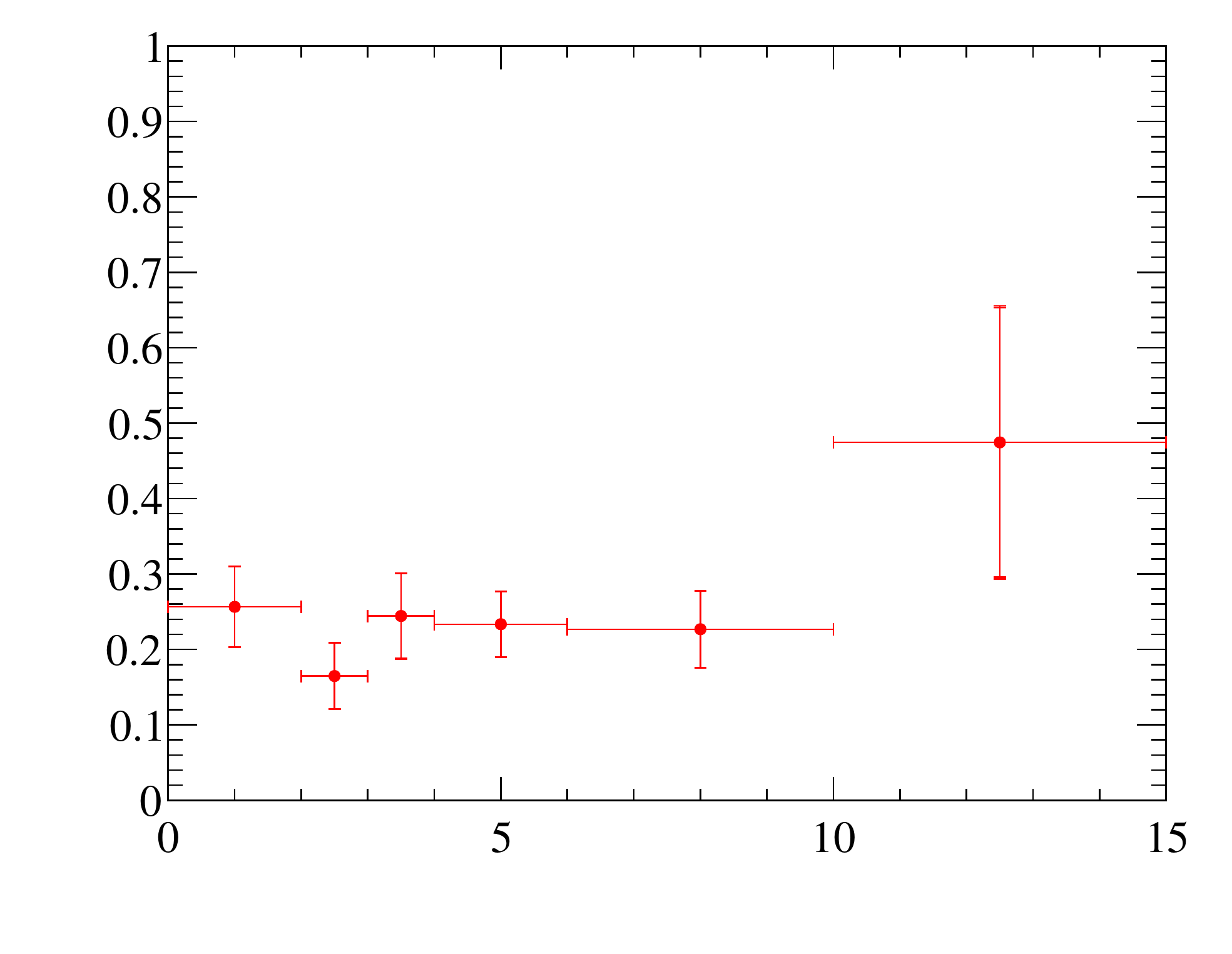}
    }
    %
    %
    \put(75, 60){
      \includegraphics*[width=75mm,height=60mm,%
      ]{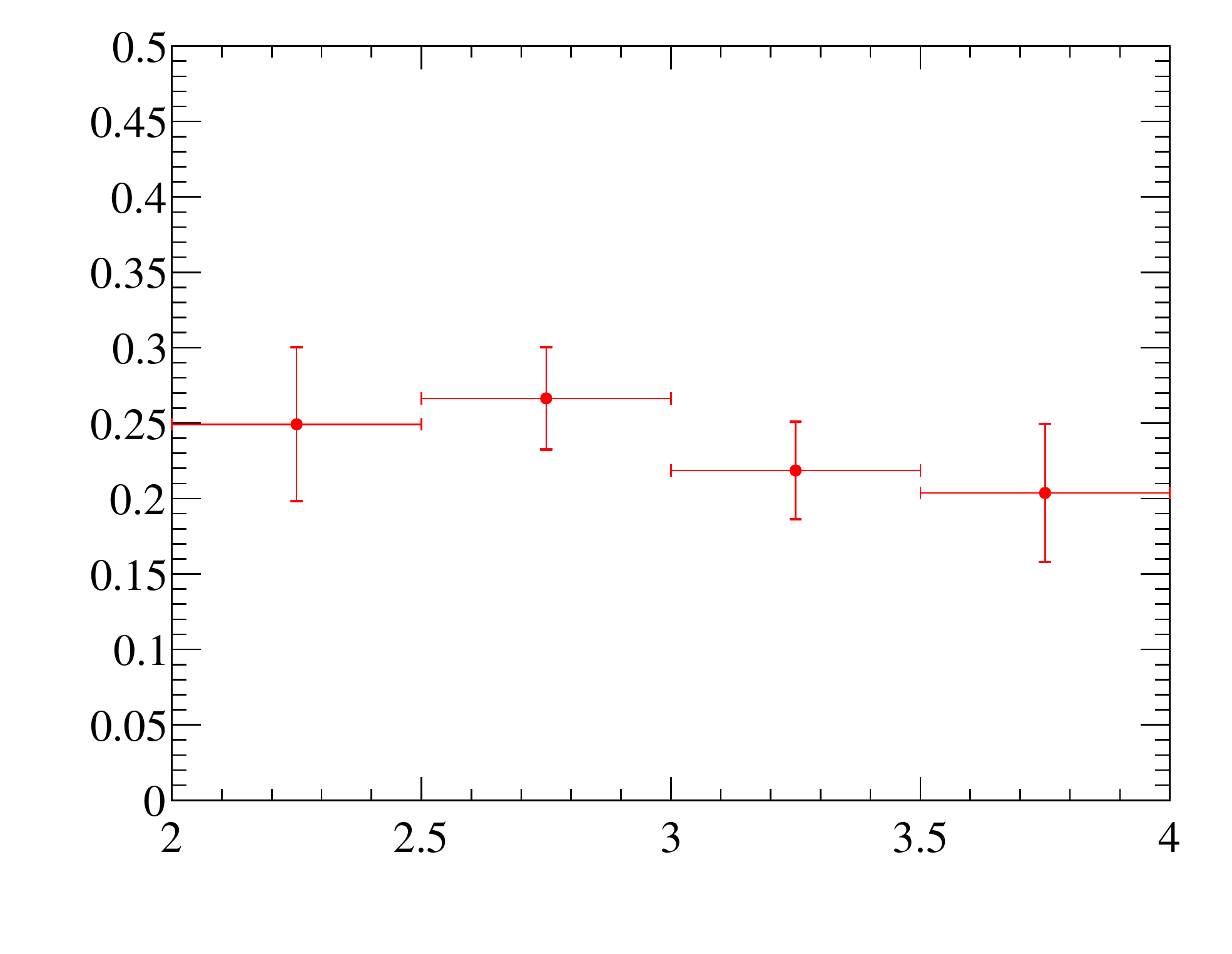}
    }
    \put( 0, 0){
      \includegraphics*[width=75mm,height=60mm,%
      ]{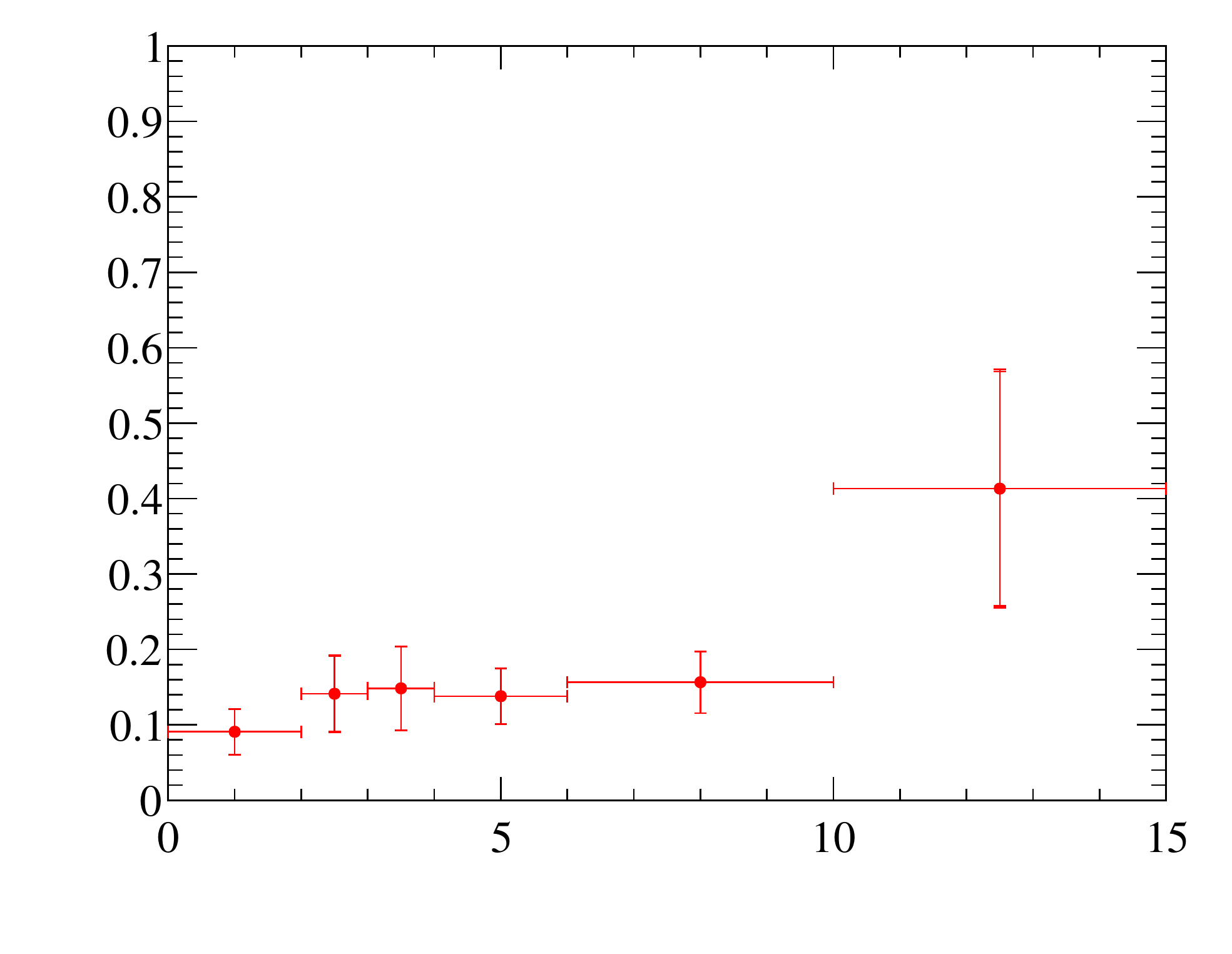}
    }
    %
    %
    \put(75, 0){
      \includegraphics*[width=75mm,height=60mm,%
      ]{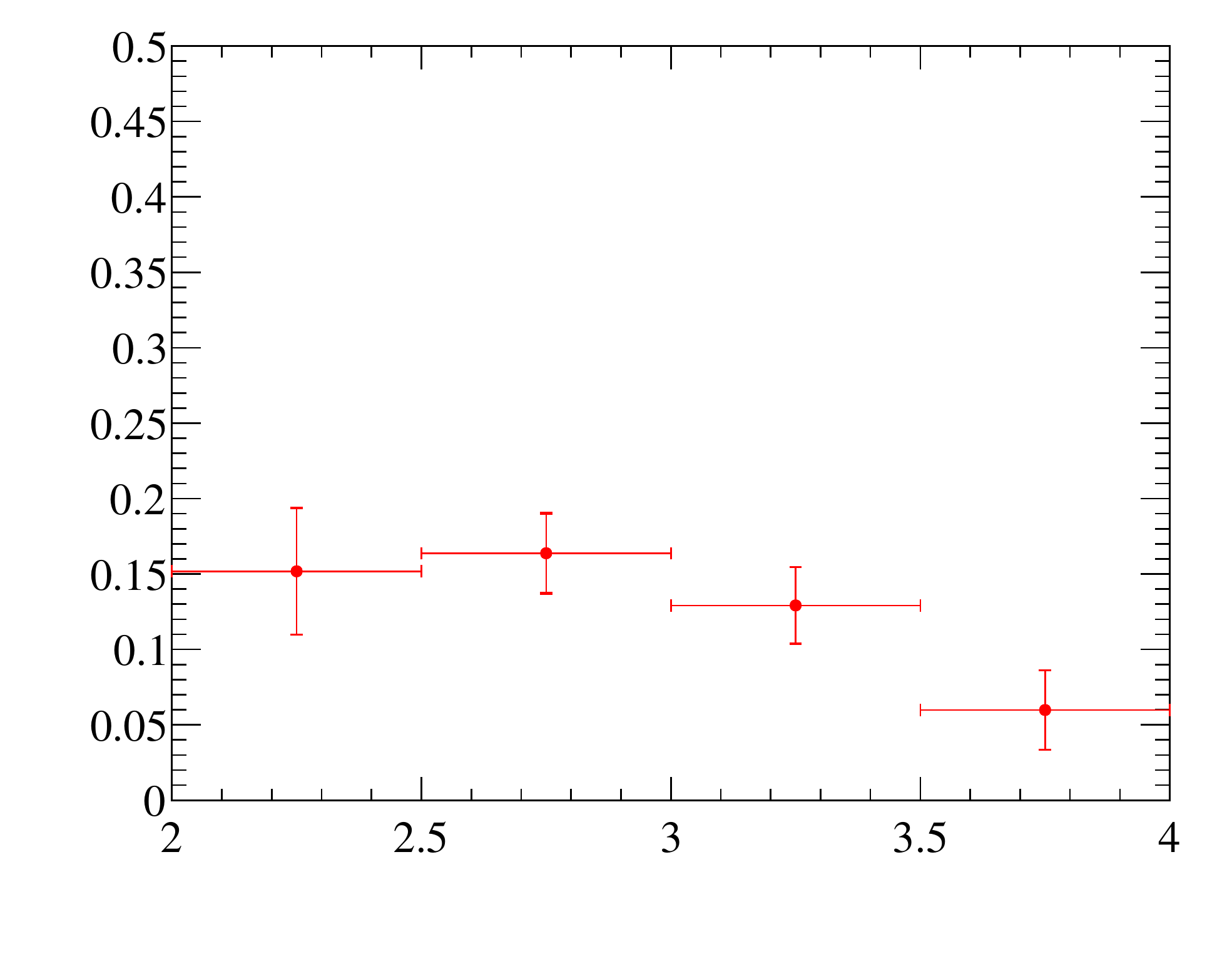}
    }
    \put( 47,110){LHCb}
    \put( 40,105){$\twos/\ones$}
    \put(122,110){LHCb}
    \put(115,105){$\twos/\ones$}
    \put( 47,50){LHCb}
    \put( 40,45){$\threes/\ones$}
    \put(122,50){LHCb}
    \put(115,45){$\threes/\ones$} 
    \put( 38, 61){ $\pt$  }
    \put( 57, 61){ $\left[\!\gevc\right]$  }
    \put(112, 61){ $y$   }
    \put( 38,  1){ $\pt$  }
    \put( 57,  1){ $\left[\!\gevc\right]$  }
    \put(112,  1){ $y$   }
    \put( 0,106) {\begin{sideways}  $\mathcal{R}^{\rm 2S/1S}$ \end{sideways}}
    \put(75,106) {\begin{sideways}  $\mathcal{R}^{\rm 2S/1S}$ \end{sideways}}
    \put( 0, 46) {\begin{sideways}  $\mathcal{R}^{\rm 3S/1S}$ \end{sideways}}
    \put(75, 46) {\begin{sideways}  $\mathcal{R}^{\rm 3S/1S}$ \end{sideways}}
  \end{picture}
  \caption {
    \small Ratios of the $\twos$ to $\ones$ and $\threes$ to $\ones$ cross-sections times
    dimuon branching fractions as functions of \pt and $y$. 
    The error bars indicate the total uncertainties 
    of the results obtained by adding statistical and systematic uncertainties in quadrature.
  }
  \label{fig:results_ratios_1}
\end{figure}
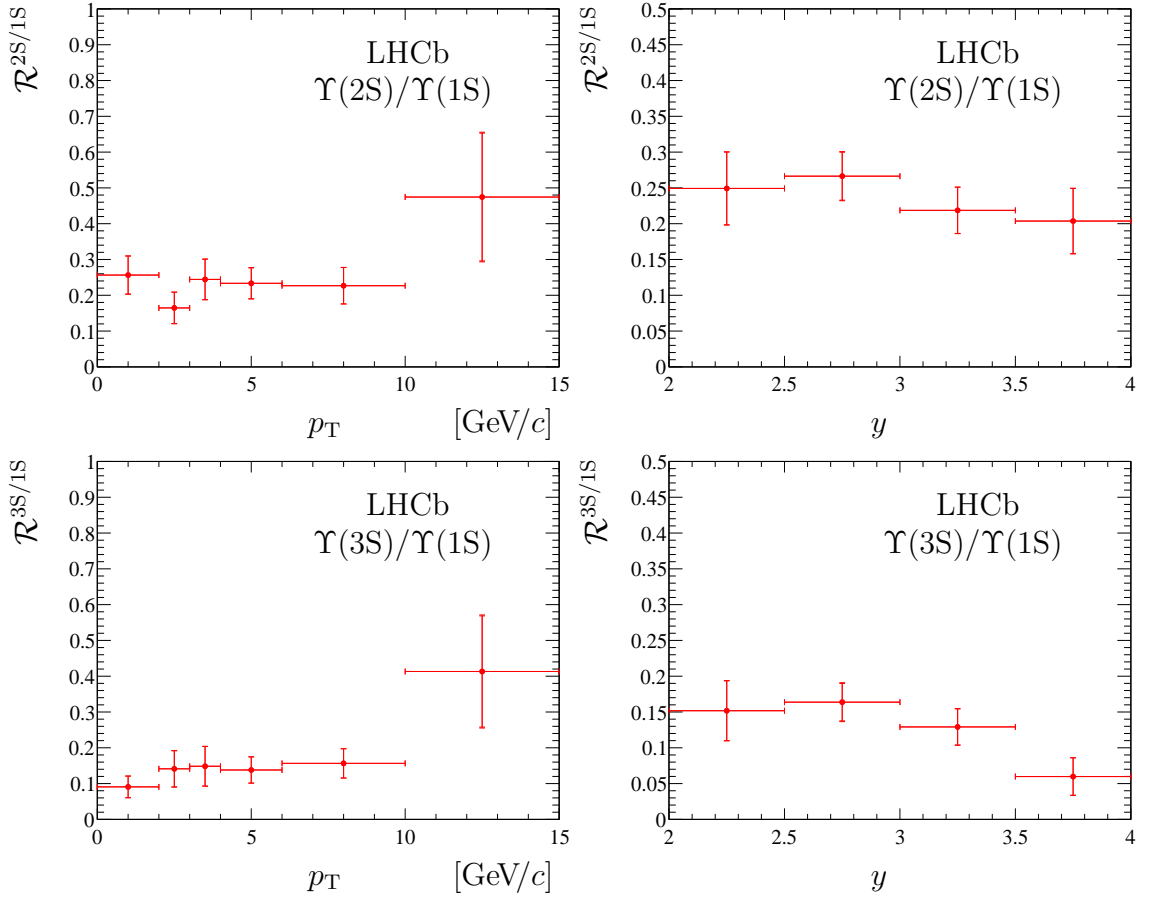

Figure~\ref{fig:results_ratios_1} illustrates the ratios of the 
$\twos$ to $\ones$, $\mathcal{R}^{\mathrm 2S/1S}$, and $\threes$ to $\ones$, 
$\mathcal{R}^{\mathrm 3S/1S}$,  cross-sections times dimuon branching 
fractions as functions of~\pt and~$y$. 
Here, most of the systematic uncertainties 
on the cross-sections cancel, 
while the statistical uncertainties remain significant. 
The ratios are found to be in good agreement with                           
the~corresponding results obtained in the previous analyses 
on \ups~production at $\sqrt{s}=7$ and $8\tev$~\cite{LHCb-PAPER-2011-036,LHCb-PAPER-2013-016}.
The measured $\mathcal{R}^{\mathrm 2S/1S}$ and 
$\mathcal{R}^{\mathrm 3S/1S}$ are also consistent with
theoretical predictions presented in 
Refs.~\cite{Kisslinger:2011fe,Kisslinger:2012tu,Kisslinger:2012np}, 
where the \threes meson is considered as a mixture of normal 
$\bquark\bquarkbar$~and hybrid $\bquark\bquarkbar\mathrm{g}$~states.
Table~\ref{tab:ratio} lists $\mathcal{R}^{\mathrm 2S/1S}$ and 
$\mathcal{R}^{\mathrm 3S/1S}$ for each \pt and $y$ bin.

To provide a reference for a future LHCb measurement of \ups~production 
with $\proton\mathrm{Pb}$~collisions at $\sqrt{s_{NN}}=5$~TeV, 
the \ups~cross-sections are measured in the reduced kinematic region
$\pt<15\gevc$ and $2.5<y<4.0$. 
The corresponding integrated cross-sections times dimuon branching fractions
in this kinematic region are
\begin{eqnarray*}
\upsigma \left( \proton\proton \rightarrow \ones   \mathrm{X} \right) \times \BR\left(\onesmm   \right) & = & 0.670 \pm 0.025  \pm 0.026\nb, \\ 
\upsigma \left( \proton\proton \rightarrow \twos   \mathrm{X} \right) \times \BR\left(\twosmm   \right) & = & 0.159 \pm 0.013  \pm 0.007\nb, \\ 
\upsigma \left( \proton\proton \rightarrow \threes \mathrm{X} \right) \times \BR\left(\threesmm \right) & = & 0.089 \pm  0.010 \pm 0.004\nb.
\end{eqnarray*}

\section{Conclusions}
\label{sec:concl}

The production of \ones, \twos and \threes mesons is observed for 
the first time in $\proton\proton$~collisions at a~centre-of-mass energy 
of $\sqrt{s}=2.76\tev$ at forward rapidities with a~data sample 
corresponding to an integrated luminosity of 3.3\invpb. 
The \ups differential production cross-sections times 
dimuon branching fractions are measured separately 
as functions of the~\ups~transverse momentum and rapidity
for $\pt<15\gevc$ and $2.0<y<4.5$. 
The~theoretical predictions, based on the next-to-leading 
order non-relativistic QCD calculation, provide a good description 
of the data at large \pt.
The ratios of the $\twos$ to $\ones$ and $\threes$ to $\ones$ 
cross-sections times dimuon branching fractions
as functions of \pt and $y$ are found to be in agreement 
with the corresponding results obtained at higher collision energies.

\section*{Acknowledgements}

\noindent 
We thank  G. Bodwin, L.~S.~Kisslinger, A.~K.~Likhoded and A.~V.~Luchinsky 
for fruitful discussions about bottomonium production. 
In addition, we are grateful to K.-T.~Chao, H.~Han and H.-S.~Shao 
for the next-to-leading order non-relativistic QCD predictions 
for prompt $\Upsilon$~production
at $\sqrt{s}=2.76\tev$. 
We also express our gratitude to our colleagues in the CERN
accelerator departments for the excellent performance of the LHC. We
thank the technical and administrative staff at the LHCb
institutes. We acknowledge support from CERN and from the national
agencies: CAPES, CNPq, FAPERJ and FINEP (Brazil); NSFC (China);
CNRS/IN2P3 and Region Auvergne (France); BMBF, DFG, HGF and MPG
(Germany); SFI (Ireland); INFN (Italy); FOM and NWO (The Netherlands);
SCSR (Poland); MEN/IFA (Romania); MinES, Rosatom, RFBR and NRC
``Kurchatov Institute'' (Russia); MinECo, XuntaGal and GENCAT (Spain);
SNSF and SER (Switzerland); NAS Ukraine (Ukraine); STFC (United
Kingdom); NSF (USA). We also acknowledge the support received from the
ERC under FP7. The Tier1 computing centres are supported by IN2P3
(France), KIT and BMBF (Germany), INFN (Italy), NWO and SURF (The
Netherlands), PIC (Spain), GridPP (United Kingdom). 
We are indebted to the communities behind the multiple open source 
software packages we depend on. We are also thankful for the computing 
resources and the access to software R\&D tools provided by Yandex LLC (Russia).

\bibliographystyle{LHCb}
\bibliography{main,LHCb-PAPER,LHCb-CONF,LHCb-DP,local}

\end{document}